\documentclass[varenna]{cimento}
\usepackage{amssymb}
\usepackage{epsfig,amsmath,graphics}
\usepackage{epstopdf}

\newcommand{\eff}{\text{eff}}
\newcommand{\keff}{k_\text{eff}}

\renewcommand{\v}[1]{\ensuremath{\mathbf{#1}}}
\newcommand{\ket}[1]{\ensuremath{\left|#1\right>}}
\newcommand{\bra}[1]{\ensuremath{\left<#1\right|}}
\newcommand{\braket}[2]{\ensuremath{\left<#1|#2\right>}}

\newcommand{\commutator}[2]{\ensuremath{\left[#1,#2\right]}}
\newcommand{\expectation}[1]{\ensuremath{\left<#1\right>}}
\newcommand{\abs}[1]{\ensuremath{\left|#1\right|}}

\def\be{\begin{equation}}
\def\ee{\end{equation}}

\newcommand{\OP}[1]{\ensuremath{\hat{#1}}}
\newcommand{\vOP}[1]{\ensuremath{\hat{\v{#1}}}}

\def\imagI{i}

\def\TotalState{\Psi}
\def\InternalState{A}
\def\ExternalState{\psi}
\def\CMState{\phi_{CM}}
\def\GalileanTrans{G}

\begin{document}


\title{Light-pulse atom interferometry}

\author{Jason M. Hogan, David M. S. Johnson \atque Mark A. Kasevich }
\institute{Department of Physics, Stanford University, Stanford,
California 94305}


\maketitle

\begin{abstract}
The light-pulse atom interferometry method is reviewed.
Applications of the method to inertial navigation and tests of the
Equivalence Principle are discussed.
\end{abstract}



\eject

\tableofcontents 

\section{Introduction}

De Broglie wave interferometry using cold atoms is emerging as a
new tool for basic science and technology.  There are numerous
approaches and applications which have evolved since the first
demonstration experiments in the early 1990's.  These notes will
not attempt an exhaustive or comprehensive survey of the field. An
excellent overview is provided in Ref. \cite{Berman} and other
lectures in this volume. These notes will focus on what has become
known as light-pulse atom interferometry, which has found fruitful
applications in gravitational physics and inertial sensor
development.

These notes are organized as follows.  We we first summarize basic
theoretical concepts.  We will then illustrate this formalism with
a discussion of applications in inertial navigation and in a
detailed design discussion of an experiment currently under
development to test the Weak Equivalence Principle.

\section{Atom interferometry overview}
By analogy with their optical counterparts, atom interferometery
seeks to exploit wave interference as a tool for precision
metrology.  In this measurement paradigm, a single particle
(photon or atom) is made to coherently propagate over two paths.
These paths are subsequently recombined using a beamsplitter, and
their relative phase becomes manifest in the probability of
detecting the particle in a given output port of the device.
Hence, measuring the particle flux at the interferometer output
ports enables determination of the phase shift.  As this relative
phase depends on physical interactions over the propagation
paths, this enables the characterization of these interactions.

A key challenge for de Broglie wave interferometers is the
development of techniques to coherently divide and recombine
atomic wavepackets. For simplicity, consider an atom with initial
momentum $\v{p}$, characterized by a wavefunction $\psi(\v{x})
\sim \exp{\left[\frac{i}{\hbar} {\bf p \cdot x}\right]}$
\footnote{In reality, the wavefunction is spatially localized, so
by the uncertainty principle, there must be a corresponding spread
in momentum about the mean value.}. The atom then is subject to a
Hamiltonian interaction which is engineered to evolve the
wavepacket into a momentum superposition state. One such
interaction spatially modulates the amplitude of the wavefunction,
so that, for example, $\exp{\left[\frac{i}{\hbar} {\bf p \cdot
x}\right]} \rightarrow f(\v{x}) \exp{\left[\frac{i}{\hbar} {\bf p
\cdot x}\right]}$, where $f(\v{x})$ is a real periodic function
with spatial frequency $\v{k}$. Fourier decomposing $f(\v{x})$
immediately shows that the final wavefunction is a coherent
superposition of momenta $\v{p}$, ${\bf p} \pm \hbar {\bf k}$,
${\bf p} \pm 2\hbar {\bf k}$, etc. In practice, such an
interaction can be implemented by passing a collimated atomic beam
through a microfabricated transmission grating, as demonstrated by
Pritchard and co-workers \cite{Pritchard}.

Another interaction is one which spatially modulates the phase of
the wavefunction.  Consider, for example, $\psi(\v{x}) \rightarrow
\exp{\left[i {\bf k \cdot x}\right]}\psi(\v{x})$. This interaction
results in a momentum translation ${\bf p} \rightarrow {\bf p}+
\hbar {\bf k}$. A particularly useful implementation of this
process imparts a spatial phase modulation by driving transitions
between internal atomic states. For simplicity, consider a
two-level atom with internal states $\ket{1}$ and $\ket{2}$ that
are resonantly coupled by an applied optical traveling wave
$\v{E}\propto\exp{\left[i {\bf k} \cdot {\bf x}\right]}$ via the
electric-dipole interaction $\vOP{\mu}\cdot\v{E}$ (where
$\vOP{\mu}$ is the dipole moment operator). If the atom is
initially prepared in state $\psi(\v{x}) \ket{1}$, then following
an interaction time $t$ its state becomes $\alpha(t) \psi(\v{x})
\ket{1}$ + $\beta(t) \exp{\left[i {\bf k} \cdot {\bf x}\right]}
\psi(\v{x}) \ket{2}$ (see Section \ref{Sec:AI Calc} for details).
The interaction time can be chosen, for example, so that
$|\alpha|$ = $|\beta|$ = $1/\sqrt{2}$ to implement a beamsplitter
(the $\pi/2$ pulse condition).  In this case, the internal state
of the atom becomes correlated with its external momentum. In
practice, two-photon stimulated Raman transitions between
groundstate hyperfine levels have proven to be particularly
fruitful for implementing this class of beamsplitter. Why?
Transitions are made between long lived hyperfine levels while the
phase grating periodicity is twice that of a single photon optical
transition (when the Raman transition is driven in a
counter-propagating beam geometry).

The above mechanisms operate in free space.  A new family of atom
optics, based on control of atom wavepacket motion in atomic
waveguides, is under development.  The basic idea is that atoms
are steered using microfabricated wires deposited on surfaces.
These are loosely analogous to optical fiber waveguides for light.
In principle, coherent beamsplitters are implemented by the
appropriate joining of waveguides.  Recently, a combination of
microwave and magnetic fields has allowed for creation of
waveguide structures capable of coherent wavefront division. These
structures have been used to demonstrate proof-of-principle
interferometer topologies which have been used to study the
coherence properties of the division process.  The notes below
will discuss free space atom optics, which have proven effective
for precision measurement of inertial forces.

Exploiting the momentum exchange principles outlined above, it
becomes straightforward to devise a de Broglie wave
interferometer which is based on sequences of light pulses.  For
example, consider a three pulse sequence based on Raman
transitions.  An initial Raman $\pi/2$ pulse places an atom in a
coherent superposition of wavepackets in states $\ket{1}$ and
$\ket{2}$ whose mean momenta differ by $\hbar \v{k}_\eff$ (here
$\v{k}_\eff$ is the effective wavevector of the Raman process,
see below). After an interrogation time $T$ these wavepackets
separate by a distance $\hbar k_\eff T/m$, where $m$ is the
atomic mass.  A subsequent optical pulse is then applied whose
duration is chosen to drive the transitions $\ket{1} \rightarrow
\ket{2}$ and $\ket{2} \rightarrow \ket{1}$ with unit probability
(a $\pi$ pulse). This pulse has the effect of redirecting the
momenta of the wavepackets so that at a time $T$ later the
wavepackets again overlap.  A final $\pi/2$ pulse then serves as
the exit beamsplitter.

\section{Phase shift determination}
\label{Sec:AI Calc}

In this section we review the method for calculating the phase
difference between the two halves of the atom at the end of the
light-pulse atom interferometer pulse sequence outlined above.
These results are well-known \cite{cct, bongs}, but we are not
aware of a complete, formal derivation of these rules in the
literature. Other equivalent formalisms for this calculation do
exist (see, for example \cite{boris, quantum_calculation}). For
Section \ref{Sec:EP phase shift calc} it is necessary to
understand the formulae for the phase difference (Section
\ref{Sec: AI answer}).  The proof of these formulae as well as a
discussion of their range of validity is given in Section
\ref{Sec:AI proof} but is not necessary for the rest of the paper.

\subsection{Phase shift formulae}
\label{Sec: AI answer}

The main result we will show is that the total phase difference
$\Delta \phi_\text{tot}$ between the two paths of an atom
interferometer may be written as the sum of three easily
calculated components:
\begin{equation}\label{Eq:DeltaPhiTotal}
\Delta \phi_\text{tot} = \Delta \phi_\text{propagation} + \Delta
\phi_\text{separation} + \Delta \phi_\text{laser}.
\end{equation}
For this calculation we take $\hbar = c = 1$.

The propagation phase $\Delta \phi_\text{propagation}$ arises from
the free--fall evolution of the atom between light pulses and is
given by \be\Delta \phi_\text{propagation} =
\sum_{\text{upper}}\left(\int_{t\!_I}^{t\!_F}\! (L_c - E_i)
dt\right) - \sum_{\text{lower}}\left(\int_{t\!_I}^{t\!_F}\! (L_c -
E_i) dt\right)\label{PropagationPhaseIntro}\ee where the sums are
over all the path segments of the upper and lower arms of the
interferometer, and $L_c$ is the classical Lagrangian evaluated
along the classical trajectory of each path segment. In addition
to the classical action, Eq. (\ref{PropagationPhaseIntro})
includes a contribution from the internal atomic energy level
$E_i$.  The initial and final times $t_I$ and $t_F$ for each path
segment, as well as $L_c$ and $E_i$, all depend on the path
segment.

The laser phase $\Delta \phi_\text{laser}$ comes from the
interaction of the atom with the laser field used to manipulate
the wavefunction at each of the beamsplitters and mirrors in the
interferometer.  At each interaction point, the component of the
state that changes momentum due to the light acquires the phase of
the laser $\phi_L(t_0,\v{x}_c(t_0)) =
\v{k}\cdot\v{x}_c(t_0)-\omega t_0+\phi$ evaluated at the classical
point of the interaction: \be\Delta \phi_\text{laser} =
\left(\sum_j
\pm\phi_L(t_j,\v{x}_u(t_j))\right)_\text{upper}-\left(\sum_j
\pm\phi_L(t_j,\v{x}_l(t_j))\right)_\text{lower}\label{LaserPhaseIntro}\ee
The sums are over all the interaction points at the times $t_j$,
and $\v{x}_u(t)$ and $\v{x}_l(t)$ are the classical trajectories
of the upper and lower arm of the interferometer, respectively.
The sign of each term depends on whether the atom gains $(+)$ or
loses $(-)$ momentum as a result of the interaction.

The separation phase $\Delta\phi_\text{separation}$ arises when
the classical trajectories of the two arms of the interferometer
do not exactly intersect at the final beamsplitter (see Fig.
\ref{Fig:SeparationPhase}).  For a separation between the upper
and lower arms of $\v{\Delta x} = \v{x}_l - \v{x}_u$, the
resulting phase shift is \be\Delta\phi_\text{separation}=
\v{\bar{p}}\cdot\v{\Delta x}\label{SeparationPhaseIntro}\ee where
$\v{\bar{p}}$ is the average classical canonical momentum of the
atom after the final beamsplitter.

\subsection{Justification of phase shift formulae}
\label{Sec:AI proof}

The interferometer calculation amounts to solving the Schrodinger
equation with the following Hamiltonian: \be \OP{H}_\text{tot} =
\OP{H}_\text{a} + \OP{H}_\text{ext} +
\OP{V}_\text{int}(\vOP{x})\ee Here $\OP{H}_\text{a}$ is the
internal atomic structure Hamiltonian, $\OP{H}_\text{ext}$ is the
Hamiltonian for the atom's external degrees of freedom (center of
mass position and momentum), and $\OP{V}_\text{int}(\vOP{x}) =
-\vOP{\mu}\cdot\v{E}(\vOP{x})$ is the atom-light interaction,
which we take to be the electric dipole Hamiltonian with
$\vOP{\mu}$ the dipole moment operator.

The calculation is naturally divided into a series of light pulses
during which $\OP{V}_\text{int}\neq 0$, and the segments between
light pulses during which $\OP{V}_\text{int} = 0$ and the atom is
in free-fall.  When the light is off, the atom's internal and
external degrees of freedom are decoupled. The internal
eigenstates satisfy \be \imagI \partial_t \ket{\InternalState_i} =
\OP{H}_\text{a} \ket{\InternalState_i} = E_i
\ket{\InternalState_i}\label{InternalStateSchrodinger}\ee and we
write the solution as $\ket{\InternalState_i} = \ket{i}e^{-\imagI
E_i (t-t_0)}$ with time-independent eigenstate \ket{i} and energy
level $E_i$.

For the external state solution $\ket{\ExternalState}$, we
initially consider $\OP{H}_\text{ext} = H(\vOP{x},\vOP{p})$ to be
an arbitrary function of the external position and momentum
operators: \be \imagI \partial_t \ket{\ExternalState} =
H(\vOP{x},\vOP{p}) \ket{\ExternalState}.
\label{ExternalStateSchrodinger}\ee  It is now useful to introduce
a Galilean transformation operator \be \OP{\GalileanTrans}_c
\equiv \OP{\GalileanTrans}(\v{x}_c,\v{p}_c,L_c) = e^{\imagI\int\!
L_c dt}  e^{-\imagI \vOP{p} \cdot \v{x}_c}  e^{\imagI
\v{p}_c\cdot\vOP{x}}\ee which consists of momentum boost by
$\v{p}_c$, a position translation by $\v{x}_c$, and a phase shift.
We choose to write \be \ket{\ExternalState} =
\OP{\GalileanTrans}_c \ket{\CMState}. \label{CMStateDef} \ee We
will show that for a large class of relevant Hamiltonians, if
$\v{x}_c$, $\v{p}_c$, and $L_c$ are taken to be the classical
position, momentum and Lagrangian, respectively, then
$\ket{\CMState}$ is a wavepacket with
$\left<\vOP{x}\right>=\left<\vOP{p}\right>=0$, and the dynamics of
$\ket{\CMState}$ do not affect the phase shift result (i.e.,
$\ket{\CMState}$ is the center of mass frame wavefunction).
However, for now we maintain generality and just treat $\v{x}_c$,
$\v{p}_c$, and $L_c$ as arbitrary functions of time.  Combining
(\ref{ExternalStateSchrodinger}) and (\ref{CMStateDef}) results in
\begin{eqnarray}
  \imagI \partial_t \ket{\CMState} &=& \left\{ \OP{\GalileanTrans}_c^\dag H(\vOP{x},\vOP{p})\OP{\GalileanTrans}_c - \imagI \OP{\GalileanTrans}_c^\dag \partial_t\OP{\GalileanTrans}_c \right\} \ket{\CMState} \\
\nonumber
  &=& \left\{ H(\vOP{x} + \v{x}_c,\vOP{p} + \v{p}_c) + \v{\dot{p}}_c\cdot\vOP{x} - \left(\vOP{p} + \v{p}_c\right)\cdot\v{\dot{x}}_c + L_c\right\} \ket{\CMState}
\end{eqnarray}
where we used the following identities:
\begin{eqnarray}
  \OP{\GalileanTrans}_c^\dag\vOP{x}\OP{\GalileanTrans}_c &=& \vOP{x} + \v{x}_c\label{GalileanIdentities}\\
  \nonumber
  \OP{\GalileanTrans}_c^\dag\vOP{p}\OP{\GalileanTrans}_c &=& \vOP{p} + \v{p}_c \\
  \nonumber
  \OP{\GalileanTrans}_c^\dag H(\vOP{x},\vOP{p})\OP{\GalileanTrans}_c &=& H(\vOP{x} + \v{x}_c,\vOP{p} + \v{p}_c)
\end{eqnarray}
Next, we Taylor expand $H(\vOP{x} + \v{x}_c,\vOP{p} + \v{p}_c)$
about $\v{x}_c$ and $\v{p}_c$, \be H(\vOP{x} + \v{x}_c,\vOP{p} +
\v{p}_c) = H(\v{x}_c,\v{p}_c) + \nabla\!_{\vOP{x}}
H(\v{x}_c,\v{p}_c)\cdot\vOP{x} + \nabla\!_{\vOP{p}}
H(\v{x}_c,\v{p}_c)\cdot\vOP{p} + \OP{H}_2\ee where $\OP{H}_2$
contains all terms that are second order or higher in $\vOP{x}$
and $\vOP{p}$.  (We will ultimately be allowed to neglect
$\OP{H}_2$ in this calculation.)  Inserting this expansion and
grouping terms yields \be\nonumber\imagI \partial_t \ket{\CMState}
= \left\{ \big(H_c - \v{\dot{x}}_c\cdot\v{p}_c + L_c\big) +
\big(\nabla\!_{\v{x}_c}H_c + \v{\dot{p}}_c\big)\cdot\vOP{x} +
\big(\nabla\!_{\v{p}_c}H_c - \v{\dot{x}}_c\big)\cdot\vOP{p} +
\OP{H}_2 \right\} \ket{\CMState}\ee where we have defined the
classical Hamiltonian $H_c \equiv H(\v{x}_c,\v{p}_c)$. If we now
let $\v{x}_c$, $\v{p}_c$, and $L_c$ satisfy Hamilton's equations,
\begin{eqnarray}
  \v{\dot{x}}_c &=& \nabla\!_{\v{p}_c}H_c \\
  \nonumber
  \v{\dot{p}}_c &=& - \nabla\!_{\v{x}_c}H_c\\
  \nonumber
  L_c &=& \v{\dot{x}}_c\cdot\v{p}_c - H_c
\end{eqnarray}
with $\v{p}_c\equiv\nabla\!_{\v{\dot{x}}_c}L_c$ the classical
canonical momentum, then $\ket{\CMState}$ must satisfy \be\imagI
\partial_t \ket{\CMState} = \OP{H}_2
\ket{\CMState}\label{CMStateSchrodinger}\ee

Next we show that it is possible to choose \ket{\CMState} with
$\expectation{\OP{x}}=\expectation{\OP{p}}=0$ for a certain class
of $\OP{H}_2$, so that $\v{x}_c$ and $\v{p}_c$ completely describe
the atom's classical center of mass trajectory.  This is known as
the semi-classical limit. Starting from Ehrenfest's theorem for
the expectation values of \ket{\CMState},
\begin{eqnarray}
  \partial_t \expectation{\OP{x}_i} &=& \imagI \expectation{\commutator{\OP{H}_2}{\OP{x}_i}} = \expectation{\partial_{\OP{p}_i}\OP{H}_2}\\
  \partial_t \expectation{\OP{p}_i} &=& \imagI \expectation{\commutator{\OP{H}_2}{\OP{p}_i}} = - \expectation{\partial_{\OP{x}_i}\OP{H}_2}
\end{eqnarray}
and expanding about $\expectation{\vOP{x}}$ and
$\expectation{\vOP{p}}$,
\begin{eqnarray}
  \nonumber\partial_t \expectation{\OP{x}_i} &=&   \left<\left.\partial_{\OP{p}_i}\OP{H}_2\right|_{\expectation{\vOP{x}},\expectation{\vOP{p}}} + \left.\partial_{\OP{p}_j}\partial_{\OP{p}_i}\OP{H}_2\right|_{\expectation{\vOP{x}},\expectation{\vOP{p}}} \left(\OP{p}_j - \expectation{\OP{p}_j}\right) + \left.\partial_{\OP{x}_j}\partial_{\OP{p}_i}\OP{H}_2\right|_{\expectation{\vOP{x}},\expectation{\vOP{p}}} \left(\OP{x}_j - \expectation{\OP{x}_j}\right) \right. \\\nonumber
  &&
  +\left.\frac{1}{2\,!}\left.\partial_{\OP{p}_i}\partial_{\OP{p}_j}\partial_{\OP{p}_k}\OP{H}_2\right|_{\expectation{\vOP{x}},\expectation{\vOP{p}}} \left(\OP{p}_j - \expectation{\OP{p}_j}\right)\left(\OP{p}_k - \expectation{\OP{p}_k}\right) + \cdots\right>\\
  \nonumber\partial_t \expectation{\OP{p}_i} &=&   \left<\left.\partial_{\OP{x}_i}\OP{H}_2\right|_{\expectation{\vOP{x}},\expectation{\vOP{p}}} + \left.\partial_{\OP{x}_j}\partial_{\OP{x}_i}\OP{H}_2\right|_{\expectation{\vOP{x}},\expectation{\vOP{p}}} \left(\OP{x}_j - \expectation{\OP{x}_j}\right) + \left.\partial_{\OP{p}_j}\partial_{\OP{x}_i}\OP{H}_2\right|_{\expectation{\vOP{x}},\expectation{\vOP{p}}} \left(\OP{p}_j - \expectation{\OP{p}_j}\right)\right. \\\nonumber
  &&+\left.\frac{1}{2\,!}\left.\partial_{\OP{x}_k}\partial_{\OP{x}_j}\partial_{\OP{x}_i}\OP{H}_2\right|_{\expectation{\vOP{x}},\expectation{\vOP{p}}} \left(\OP{x}_j - \expectation{\OP{x}_j}\right)\left(\OP{x}_k - \expectation{\OP{x }_k}\right) + \cdots\right>
\end{eqnarray}
we find the following:
\begin{eqnarray}
  \partial_t \expectation{\OP{x}_i} &=& \left.\partial_{\OP{p}_i}\OP{H}_2\right|_{\expectation{\vOP{x}},\expectation{\vOP{p}}} +
  \frac{1}{2\,!}\left.\partial_{\OP{p}_k}\partial_{\OP{p}_j}\partial_{\OP{p}_i}\OP{H}_2\right|_{\expectation{\vOP{x}},\expectation{\vOP{p}}} \Delta p_{jk}^2 + \cdots \label{EhrenfestX}\\
  \partial_t \expectation{\OP{p}_i} &=& -\left.\partial_{\OP{x}_i}\OP{H}_2\right|_{\expectation{\vOP{x}},\expectation{\vOP{p}}} -
  \frac{1}{2\,!}\left.\partial_{\OP{x}_k}\partial_{\OP{x}_j}\partial_{\OP{x}_i}\OP{H}_2\right|_{\expectation{\vOP{x}},\expectation{\vOP{p}}} \Delta x_{jk}^2 + \cdots\label{EhrenfestP}
\end{eqnarray}
where $\Delta
x_{jk}^2\equiv\expectation{\OP{x}_j\OP{x}_k}-\expectation{\OP{x}_j}\expectation{\OP{x}_k}$
and $\Delta
p_{jk}^2\equiv\expectation{\OP{p}_j\OP{p}_k}-\expectation{\OP{p}_j}\expectation{\OP{p}_k}$
are measures of the wavepacket's width in phase space \footnote{In
general, there will also be cross terms with phase space width
such as $\expectation{\OP{x}_j\OP{p}_k} -
\expectation{\OP{x}_j}\expectation{\OP{p}_k}$.}.  This shows that
if $\OP{H}_2$ contains no terms higher than second order in
$\vOP{x}$ and $\vOP{p}$, then Ehrenfest's theorem reduces to
Hamilton's equations, and the expectation values follow the
classical trajectories.  Furthermore, this implies that we can
choose \ket{\CMState} to be the wavefunction in the atom's rest
frame, since $\expectation{\OP{x}}=\expectation{\OP{p}}=0$ is a
valid solution to Eqs. (\ref{EhrenfestX}) and (\ref{EhrenfestP})
so long as all derivatives of $\OP{H}_2$ higher than second order
vanish.  In addition, even when this condition is not strictly
met, it is often possible to ignore the non-classical corrections
to the trajectory so long as the phase space widths $\Delta
x_{jk}$ and $\Delta p_{jk}$ are small compared to the relevant
derivatives of $\OP{H}_2$ (i.e., the semi-classical
approximation).  For example, such corrections are present for an
atom propagating in the non-uniform gravitational field $g$ of the
Earth for which
$\partial_{\OP{r}}\partial_{\OP{r}}\partial_{\OP{r}}\OP{H}_2\sim
\partial^2_r g$.  Assuming an atom wavepacket width $\Delta
x\lesssim 1~\text{mm}$, the deviation from the classical
trajectory is $\partial_t \expectation{\OP{p}}\sim (\partial^2_r
g)\Delta x^2 \lesssim 10^{-20}g$, which is a negligibly small
correction even in the context of the $\sim 10^{-15}g$ apparatus
we describe below for testing the Equivalence Principle.

The complete solution for the external wavefunction requires a
solution of Eq. (\ref{CMStateSchrodinger}) for $\ket{\CMState}$,
but this is non-trivial for general $\OP{H}_2$.  In the simplified
case where $\OP{H}_2$ is second order in $\vOP{x}$ and $\vOP{p}$,
the exact expression for the propagator is known
\cite{ExactPropagator} and may be used to determine the phase
acquired by $\ket{\CMState}$.  However, this step is not necessary
for our purpose, because for second order external Hamiltonians
the operator $\OP{H}_2$ does not depend on either $\v{x}_c$ or
$\v{p}_c$.  In this restricted case, the solution for the rest
frame wavefunction $\ket{\CMState}$ does not depend on the atom's
trajectory.  Therefore, any additional phase evolution in
$\ket{\CMState}$ must be the same for both arms of the
interferometer and so does not contribute to the phase difference.
This argument breaks down for more general $\OP{H}_2$, as does the
semi-classical description of the atom's motion, but the
corrections will depend on the width of $\ket{\CMState}$ in phase
space as shown in Eqs. (\ref{EhrenfestX}) and (\ref{EhrenfestP}).
We ignore all such wavepacket-structure induced phase shifts in
this analysis by assuming that the relevant moments $\{\Delta
x_{jk},\Delta p_{jk},\ldots\}$ are sufficiently small so that
these corrections can be neglected.  As shown above for the
non-uniform ($\partial^2_r g \neq 0$) gravitational field of the
Earth, this condition is easily met in many experimentally
relevant situations.

Finally, we can write the complete solution for the free
propagation between the light pulses:
\be\braket{\v{x}}{\ExternalState,\InternalState_i}=\bra{\v{x}}\OP{\GalileanTrans}_c\ket{\CMState}\ket{\InternalState_i}
= e^{\imagI\int_{t\!_I}^{t\!_F}\! L_c dt}  e^{\imagI
\v{p}_c\cdot\left(\v{x} - \v{x}_c \right)}\CMState\!\left(\v{x} -
\v{x}_c \right)\ket{i}e^{-\imagI E_i (t_F-t_I)}\ee We see that
this result takes the form of a traveling wave with de Broglie
wavelength set by $\v{p}_c$ multiplied by an envelope function
$\CMState(\v{x})$, both of which move along the classical path
$\v{x}_c$. Also, the wavepacket accumulates a propagation phase
shift given by the classical action along this path, as well as an
additional phase shift arising from the internal atomic energy:
\be\Delta \phi_\text{propagation} =
\sum_{\text{upper}}\left(\int_{t\!_I}^{t\!_F}\! (L_c - E_i)
dt\right) - \sum_{\text{lower}}\left(\int_{t\!_I}^{t\!_F}\! (L_c -
E_i) dt\right)\label{PropagationPhase}\ee where the sums are over
all the path segments of the upper and lower arms of the
interferometer, and $t_I$, $t_F$, $L_c$, and $E_i$ all depend on
the path.

Next, we consider the time evolution while the light is on and
$\OP{V}_\text{int}\neq 0$.  In this case, the atom's internal and
external degrees of freedom are coupled by the electric dipole
interaction, so we work in the interaction picture using the
following state ansatz:
\be\ket{\TotalState}=\int\!d\v{p}\sum_i\,c_i(\v{p})\ket{\ExternalState_\v{p}}\ket{\InternalState_i}\label{InteractionPictureStateAnsatz}\ee
where we have used the momentum space representation of
\ket{\CMState} and so
$\ket{\ExternalState_\v{p}}\equiv\OP{\GalileanTrans}_c e^{-\imagI
\OP{H}_2(t-t_0)}\ket{\v{p}}$. Inserting this state into the
Schrodinger equation gives the interaction picture equations,
\begin{eqnarray}
  \imagI \partial_t \ket{\TotalState} &=& \imagI \int\!d\v{p}\sum_i\,\frac{\partial c_i(\v{p})}{\partial t}\ket{\ExternalState_\v{p}}\ket{\InternalState_i} + \OP{H}_\text{a}\ket{\TotalState}+\OP{H}_\text{ext}\ket{\TotalState}= \OP{H}_\text{tot}\ket{\TotalState}\\
  \Rightarrow \dot{c}_i(\v{p}) &\equiv& \frac{\partial c_i(\v{p})}{\partial t} = \frac{1}{\imagI} \int\!d\v{p'}\sum_j\,c_j(\v{p'})\bra{\InternalState_i}\bra{\ExternalState_\v{p}}\OP{V}_\text{int}(\vOP{x})\ket{\ExternalState_\v{p'}}\ket{\InternalState_j}\label{InteractionPictureEqs1}
\end{eqnarray}
where we used (\ref{InternalStateSchrodinger}) and
(\ref{ExternalStateSchrodinger}) as well as the orthonormality of
$\ket{\InternalState_i}$ and $\ket{\ExternalState_\v{p}}$.  The
interaction matrix element can be further simplified by
substituting in \ket{\ExternalState_\v{p}} and using identity
(\ref{GalileanIdentities}):
\begin{eqnarray}
\bra{\ExternalState_\v{p}}\OP{V}_\text{int}(\vOP{x})\ket{\ExternalState_\v{p'}}
&=& \bra{\v{p}}e^{\imagI \OP{H}_2(t-t_0)}
\OP{V}_\text{int}(\vOP{x}+\v{x}_c) e^{-\imagI
\OP{H}_2(t-t_0)}\ket{\v{p'}} \\\nonumber &=&
\bra{\v{p}}\OP{V}_\text{int}(\vOP{x}+\v{x}_c) \ket{\v{p'}}
e^{\imagI\left(\frac{\v{p}^2}{2m}-\frac{\v{p'}^2}{2m}\right)(t-t_0)}
\end{eqnarray}
where we have made the simplifying approximation that $\OP{H}_2
\approx \frac{\vOP{p}^2}{2m}$.  This approximation works well as
long as the light pulse time $\tau\equiv t-t_0$ is short compared
to the time scale associated with the terms dropped from
$\OP{H}_2$.  For example, for an atom in the gravitational field
of Earth, this approximation ignores the contribution $m
(\partial_r g)\OP{x}^2$ from the gravity gradient, which for an
atom of size $\Delta x \approx 1~\text{mm}$ leads to a frequency
shift $\sim m (\partial_r g)\Delta x^2\sim 1~\text{mHz}$.  For a
typical pulse time $\tau < 1~\text{ms}$, the resulting errors are
$\lesssim 1~\mu\text{rad}$ and can usually be neglected.
Generally, in this analysis we will assume the short pulse (small
$\tau$) limit and ignore all effects that depend on the finite
length of the light pulse.  These systematic effects can sometimes
be important, but they are calculated
elsewhere\cite{AntoineFiniteTime}\cite{JansenThesis}.  In the case
of the $^{87}$\!Rb--$^{85}$\!Rb Equivalence Principle experiment
we discuss below, such errors are common-mode suppressed in the
differential signal because we use the same laser pulse to
manipulate both atoms simultaneously.

As mentioned before, we typically use a two photon process for the
atom optics (i.e., Raman or Bragg) in order to avoid transferring
population to the short-lived excited state. However, from the
point of view of the current analysis, these three-level systems
can typically be reduced to effective two-level
systems\cite{MolerRaman}\cite{ShoreBragg}. Since the resulting
phase shift rules are identical,  we will assume a two-level atom
coupled to a single laser frequency to simplify the analysis.
Assuming a single traveling wave excitation
$\v{E}(\vOP{x})=\v{E}_0 \cos{\left(\v{k}\cdot\vOP{x}-\omega
t+\phi\right)}$, Eq. (\ref{InteractionPictureEqs1}) becomes \be
\dot{c}_i(\v{p}) = \frac{1}{2\imagI}
\int\!d\v{p'}\sum_j\,\Omega_{ij}\,c_j(\v{p'})
\bra{\v{p}}\left(e^{\imagI\left(\v{k}\cdot(\vOP{x}+\v{x}_c)-\omega
t+\phi\right)}+h.c.\right)\ket{\v{p'}}e^{\imagI\int_{t\!_0}^{t}\!\omega_{ij}+\frac{\v{p}^2}{2m}-\frac{\v{p'}^2}{2m}dt}
\label{InteractionPictureEqs2}\ee
where the Rabi frequency is defined as $\Omega_{ij}\equiv\bra{i}(-\vOP{\mu}\cdot\v{E}_0)\ket{j}$ and $\omega_{ij}\equiv E_i - E_j$. Now we insert the identity         
\be\v{k}\cdot(\vOP{x}+\v{x}_c)-\omega t + \phi=
\underbrace{~\bigg{.}\v{k}\cdot\vOP{x}~}_\text{boost} +
\underbrace{\bigg{.}\Big{(}\v{k}\cdot\v{x}_c(t_0)-\omega
t_0+\phi\Big{)}}_\text{laser
phase}+\underbrace{\int_{t_0}^{t}\!(\v{k}\cdot\v{\dot{x}}_c-\omega)dt}_\text{Doppler
shift}\ee into Eq. (\ref{InteractionPictureEqs2}) and perform the
integration over $\v{p'}$ using
$\bra{\v{p}}e^{\pm\imagI\v{k}\cdot\vOP{x}}\ket{\v{p'}}=\braket{\v{p}}{\v{p'}\pm
\v{k}}$:
\begin{eqnarray}
\dot{c}_i(\v{p}) =
\frac{1}{2\imagI}\sum_j\,\Omega_{ij}\,\left\{c_j(\v{p}-\v{k})e^{\imagI
\phi_L}
e^{\imagI\int_{t\!_0}^{t}\!(\omega_{ij}-\omega+\v{k}\cdot\v{\dot{x}}_c
+ \frac{\v{k}\cdot\v{p}}{m}-\frac{\v{k}^2}{2m})dt}
+\right.\\\nonumber \left. c_j(\v{p}+\v{k})e^{-\imagI \phi_L}
e^{-\imagI\int_{t\!_0}^{t}\!(-\omega_{ij}-\omega+\v{k}\cdot\v{\dot{x}}_c
+ \frac{\v{k}\cdot\v{p}}{m}+\frac{\v{k}^2}{2m})dt}\right\}
\label{InteractionPictureEqs3}\end{eqnarray} where we define the
laser phase at point $\{t_0,\v{x}_c(t_0)\}$ as $\phi_L\equiv
\v{k}\cdot\v{x}_c(t_0)-\omega t_0+\phi$. Finally, we impose the
two-level constraint $(i=1,2)$ and consider the coupling between
$c_1(\v{p})$ and $c_2(\v{p}+\v{k})$:
\begin{eqnarray}
  \dot{c}_1(\v{p}) &=& \frac{1}{2\imagI}\Omega\,c_2(\v{p}+\v{k})e^{-\imagI \phi_L} e^{-\imagI\int_{t\!_0}^{t}\!\Delta(\v{p})dt}\label{InteractionPictureEqs4}\\\nonumber
  \dot{c}_2(\v{p}+\v{k}) &=& \frac{1}{2\imagI}\Omega^\ast c_1(\v{p})e^{\imagI \phi_L} e^{\imagI\int_{t\!_0}^{t}\!\Delta(\v{p})dt}
\end{eqnarray}
Here the detuning is $\Delta(\v{p})\equiv
\omega_0-\omega+\v{k}\cdot(\v{\dot{x}}_c +
\frac{\v{p}}{m})+\frac{\v{k}^2}{2m}$, the Rabi frequency is
$\Omega \equiv \Omega_{12} = (\Omega_{21})^\ast$, and
$\omega_0\equiv \omega_{21}>0$.  In arriving at Eqs.
(\ref{InteractionPictureEqs4}) we made the rotating wave
approximation\cite{AllenEberly}, dropping terms that oscillate at
$(\omega_0+\omega)$ compared to those oscillating at
$(\omega_0-\omega)$. Also, $\Omega_{ii}=0$ since the
$\ket{\InternalState_i}$ are eigenstates of parity and $\vOP{\mu}$
is odd.

The general solution to (\ref{InteractionPictureEqs4}) is \be
\qquad\left(\!\!\!\begin{array}{c}
        c_1(\v{p},t) \\
        c_2(\v{p}+\v{k},t)
      \end{array}
\!\!\!\right)=\left(\!\!\!
                \begin{array}{cc}
                  \Lambda_c(\v{p})e^{-\frac{\imagI}{2}\Delta(\v{p})\tau} &\! -\imagI\Lambda_s(\v{p})e^{-\frac{\imagI}{2}\Delta(\v{p})\tau}e^{-\imagI\phi_L} \\
                  -\imagI\Lambda_s^\ast(\v{p})e^{\frac{\imagI}{2}\Delta(\v{p})\tau}e^{\imagI\phi_L} &\! \Lambda_c^\ast(\v{p})e^{\frac{\imagI}{2}\Delta(\v{p})\tau} \\
                \end{array}
              \!\!\!\right)
\left(\!\!\!\begin{array}{c}
        c_1(\v{p},t_0) \\
        c_2(\v{p}+\v{k},t_0)
      \end{array}
\!\!\!\right) \ee
\begin{eqnarray}
  \qquad\qquad\Lambda_c(\v{p}) &=& \cos{\left(\frac{1}{2}\sqrt{\Delta(\v{p})^2+\left|\Omega\right|^2}\,\tau\right)} + \imagI\frac{\Delta(\v{p})}{\sqrt{\Delta(\v{p})^2+\left|\Omega\right|^2}}\sin{\left(\frac{1}{2}\sqrt{\Delta(\v{p})^2+\left|\Omega\right|^2}\,\tau\right)}\\
  \qquad\qquad\Lambda_s(\v{p}) &=& \frac{\Omega}{\sqrt{\Delta(\v{p})^2+\left|\Omega\right|^2}}\sin{\left(\frac{1}{2}\sqrt{\Delta(\v{p})^2+\left|\Omega\right|^2}\,\tau\right)}
\end{eqnarray}
In integrating (\ref{InteractionPictureEqs4}) we applied the short
pulse limit in the sense of $\v{k}\cdot\v{\ddot{x}}_c \tau^2\ll
1$, ignoring changes of the atom's velocity during the pulse.  For
an atom falling in the gravitational field of the Earth, even for
pulse times $\tau \sim 10~\mu\text{s}$ this term is $\sim k g
\tau^2\sim 10^{-2}~\text{rad}$ which is non-negligible at our
level of required precision.  However, for pedagogical reasons we
ignore this error here.  Corrections due to the finite pulse time
are suppressed in the proposed differential measurement between Rb
isotopes since we use the same laser to simultaneously manipulate
both species (see Section \ref{Sec: EP experiment overview}).

For simplicity, from now on we assume the light pulses are on
resonance: $\Delta(0)=0$.  We also take the short pulse limit in
the sense of $\left|\Delta(\v{p})-\Delta(0)\right|\tau\ll 1$ so
that we can ignore all detuning systematics.  This condition is
automatically satisfied experimentally, since only the momentum
states that fall within the Doppler width $\sim \tau^{-1}$ of the
pulse will interact efficiently with the light. \be
\left(\!\!\!\begin{array}{c}
        c_1(\v{p},t) \\
        c_2(\v{p}+\v{k},t)
      \end{array}
\!\!\!\right)=\left(\!\!\!
                \begin{array}{cc}
                  \Lambda_c & -\imagI\Lambda_s e^{-\imagI\phi_L} \\
                  -\imagI\Lambda_s^\ast e^{\imagI\phi_L} & \Lambda_c \\
                \end{array}
              \!\!\!\right)
\left(\!\!\!\begin{array}{c}
        c_1(\v{p},t_0) \\
        c_2(\v{p}+\v{k},t_0)
      \end{array}
\!\!\!\right)\qquad \begin{array}{c}
        \Lambda_c = \cos{\frac{\left|\Omega\right| \tau}{2}}\\
        \Lambda_s = \frac{\Omega}{\left|\Omega\right|}\sin{\frac{\left|\Omega\right| \tau}{2}}
      \end{array}
      \label{OnResonanceBeamsplitterMatrix}
\ee In the case of a beamsplitter ($\frac{\pi}{2}$ pulse), we
choose $\left|\Omega\right| \tau=\frac{\pi}{2}$, whereas for a
mirror ($\pi$ pulse) we set $\left|\Omega\right| \tau=\pi$:
\be\Lambda_{\pi/2}=\left(\!\!\!
                             \begin{array}{cc}
                               \frac{1}{\sqrt{2}} & \frac{-\imagI}{\sqrt{2}}\,e^{-\imagI\phi_L} \\
                               \frac{-\imagI}{\sqrt{2}}\,e^{\imagI\phi_L} & \frac{1}{\sqrt{2}} \\
                             \end{array}
                           \!\!\!\right)
\qquad \Lambda_{\pi}=\left(\!\!\!
                             \begin{array}{cc}
                               0 & -\imagI\,e^{-\imagI\phi_L} \\
                               -\imagI\,e^{\imagI\phi_L} & 0 \\
                             \end{array}
                           \!\!\!\right)\ee
These matrices encode the rules for the imprinting of the laser's
phase on the atom: the component of the atom $c_1(\v{p},t_0)$ that
gains momentum from the light (absorbs a photon) picks up a phase
$+\phi_L$, and the component of the atom $c_2(\v{p}+\v{k},t_0)$
that loses momentum to the light (emits a photon) picks up a phase
$-\phi_L$.  Symbolically,
\begin{eqnarray}
  \ket{\v{p}} &\rightarrow& \ket{\v{p}+\v{k}}e^{\imagI\phi_L} \label{laserPhaseRule1}\\
  \ket{\v{p}+\v{k}} &\rightarrow& \ket{\v{p}}e^{-\imagI\phi_L} \label{laserPhaseRule2}
\end{eqnarray}
As a result, the total laser phase shift is \be\Delta
\phi_\text{laser} = \left(\sum_j
\pm\phi_L(t_j,\v{x}_u(t_j))\right)_\text{upper}-\left(\sum_j
\pm\phi_L(t_j,\v{x}_l(t_j))\right)_\text{lower}\label{LaserPhase}\ee
where the sums are over all of the atom--laser interaction points
$\{t_j,\v{x}_u(t_j)\}$ and $\{t_j,\v{x}_l(t_j)\}$ along the upper
and lower arms, respectively, and the sign is determined by Eqs.
(\ref{laserPhaseRule1}) and (\ref{laserPhaseRule2}).

\begin{figure}
\begin{center}
\includegraphics[width=300pt]{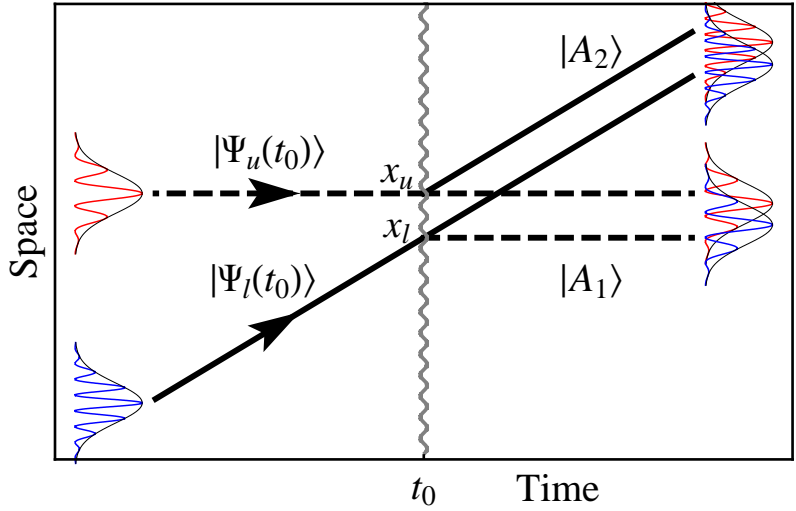}
\caption{ \label{Fig:SeparationPhase} Separation Phase. This is a
magnified view of the end of the interferometer which shows the
upper and lower arms converging at the final beamsplitter at time
$t_0$, and the resulting interference. The dashed and solid lines
designate the components of the wavefunction in internal states
$\ket{\InternalState_1}$ and $\ket{\InternalState_2}$,
respectively. After the beamsplitter, each output port consists of
a superposition of wavepackets from the upper and lower arm. Any
offset $\v{\Delta x}= \v{x}_l - \v{x}_u$ between the centers of
the wavepacket contributions to a given output port results in a
separation phase shift.}
\end{center}
\end{figure}

The final contribution to $\Delta \phi_\text{tot}$ is the
separation phase, $\Delta \phi_\text{separation}$. As shown in
Fig. \ref{Fig:SeparationPhase}, this shift arises because the
endpoints of the two arms of the interferometer need not coincide
at the time of the final beamsplitter. To derive the expression
for separation phase, we write the state of the atom at time
$t=t_0+\tau$ just after the final beamsplitter pulse as
\be\ket{\TotalState(t)}=\ket{\TotalState_u(t)}+\ket{\TotalState_l(t)}\ee
where $\ket{\TotalState_u(t)}$ and $\ket{\TotalState_l(t)}$ are
the components of the final state that originate from the upper
and lower arms, respectively. Just before the final beamsplitter
pulse is applied, we write the state of each arm as
\begin{eqnarray}
  \ket{\TotalState_u(t_0)} &=& \int\!d\v{p}\,c_1(\v{p},t_0)\OP{\GalileanTrans}_u(t_0)\ket{\v{p}}\ket{\InternalState_1}e^{\imagI \theta_u} \label{IC-UpperState}\\
  \ket{\TotalState_l(t_0)} &=& \int\!d\v{p}\,c_2(\v{p},t_0)\OP{\GalileanTrans}_l(t_0)\ket{\v{p}}\ket{\InternalState_2}e^{\imagI \theta_l}\label{IC-LowerState}
\end{eqnarray}
where $\OP{\GalileanTrans}_u\equiv
\OP{\GalileanTrans}(\v{x}_u,\v{p}_u,L_u)$ and
$\OP{\GalileanTrans}_l\equiv
\OP{\GalileanTrans}(\v{x}_l,\v{p}_l,L_l)$ are the Galilean
transformation operators for the upper and lower arm,
respectively.  These operators translate each wavepacket in phase
space to the appropriate position ($\v{x}_u$ or $\v{x}_l$) and
momentum ($\v{p}_u$ or $\v{p}_l$).  Here we have assumed for
clarity that prior to the final beamsplitter the upper and lower
arms are in internal states $\ket{\InternalState_1}$ and
$\ket{\InternalState_2}$ with amplitudes $c_1(\v{p},t_0)$ and
$c_2(\v{p},t_0)$, respectively; identical results are obtained in
the reversed case.  We have also explicitly factored out the
dynamical phases $\theta_u$ and $\theta_l$ accumulated along the
upper and lower arms, respectively, which contain by definition
all contributions to laser phase and propagation phase acquired
prior to the final beamsplitter.

We write the wavefunction components after the beamsplitter in the
form of Eq. (\ref{InteractionPictureStateAnsatz}):
\begin{eqnarray}
  \ket{\TotalState_u(t)} &=& \int\!d\v{p}\sum_i\,c_i^{(u)}(\v{p},t)\OP{\GalileanTrans}_u\ket{\v{p}}\ket{\InternalState_i} \\
  \ket{\TotalState_l(t)} &=& \int\!d\v{p}\sum_i\,c_i^{(l)}(\v{p},t)\OP{\GalileanTrans}_l\ket{\v{p}}\ket{\InternalState_i}
\end{eqnarray}
where we invoked the short pulse limit so that
$e^{-\imagI\OP{H}_2\tau}\approx 1$.  Next we time evolve the
states using Eq. (\ref{OnResonanceBeamsplitterMatrix}) assuming a
perfect $\frac{\pi}{2}$ pulse and using the initial conditions
given in Eqs. (\ref{IC-UpperState}) and (\ref{IC-LowerState}):
namely, $c_1^{(u)}(\v{p},t_0)=c_1(\v{p},t_0)e^{\imagI \theta_u}$
and $c_2^{(u)}(\v{p},t_0)=0$ for the upper arm and
$c_1^{(l)}(\v{p},t_0)=0$ and
$c_2^{(l)}(\v{p},t_0)=c_2(\v{p},t_0)e^{\imagI \theta_l}$ for the
lower arm.
\begin{eqnarray}
  \qquad\qquad\ket{\TotalState_u(t)} &=& \int\!d\v{p}\,c_1(\v{p},t_0)\left\{ \frac{1}{\sqrt{2}}\OP{\GalileanTrans}_u\ket{\v{p}}\ket{\InternalState_1}+\frac{-\imagI}{\sqrt{2}}e^{\imagI\phi_L(\v{x}_u)}\OP{\GalileanTrans}_u\ket{\v{p}+\v{k}}\ket{\InternalState_2} \right\}e^{\imagI \theta_u} \\
  \qquad\qquad\ket{\TotalState_l(t)} &=& \int\!d\v{p}\,c_2(\v{p}+\v{k},t_0)\left\{ \frac{-\imagI}{\sqrt{2}}e^{-\imagI\phi_L(\v{x}_l)}\OP{\GalileanTrans}_l\ket{\v{p}}\ket{\InternalState_1}+\frac{1}{\sqrt{2}}\OP{\GalileanTrans}_l\ket{\v{p}+\v{k}}\ket{\InternalState_2} \right\}e^{\imagI \theta_l}
\end{eqnarray}
We now project into position space and perform the $\v{p}$
integrals,
\begin{eqnarray}
  \qquad\qquad\braket{\v{x}}{\TotalState_u(t)} &=& \frac{c_1(\v{x}-\v{x}_u,t_0)}{\sqrt{2}}\left\{ e^{\imagI\v{p}_u\cdot(\v{x}-\v{x}_u)}\ket{\InternalState_1}-\imagI e^{\imagI\phi_L(\v{x}_u)}e^{\imagI(\v{p}_u+\v{k})\cdot(\v{x}-\v{x}_u)}\ket{\InternalState_2} \right\}e^{\imagI \theta_u} \\
  \qquad\qquad\braket{\v{x}}{\TotalState_l(t)} &=& \frac{c_2(\v{x}-\v{x}_l,t_0)}{\sqrt{2}}\left\{-\imagI e^{-\imagI\phi_L(\v{x}_l)}e^{\imagI(\v{p}_l-\v{k})\cdot(\v{x}-\v{x}_l)}\ket{\InternalState_1}+ e^{\imagI\v{p}_l\cdot(\v{x}-\v{x}_l)}\ket{\InternalState_2} \right\}e^{\imagI \theta_l}
\end{eqnarray}
where we identified the Fourier transformed amplitudes using
$c_i(\v{x}-\v{x}_c,
t_0)=\int\!d\v{p}\,\braket{\v{x}-\v{x}_c}{\v{p}}c_i(\v{p},t_0)$.
The resulting interference pattern in position space is therefore
\begin{eqnarray}
\nonumber\braket{\v{x}}{\TotalState(t)} &=&
\braket{\v{x}}{\TotalState_u(t)}+\braket{\v{x}}{\TotalState_l(t)}\\\nonumber
&=& \frac{1}{\sqrt{2}}\ket{\InternalState_1}\left\{
c_1(\v{x}-\v{x}_u,t_0)e^{\imagI
\theta_u}e^{\imagI\v{p}_u\cdot(\v{x}-\v{x}_u)} -
\imagI\,c_2(\v{x}-\v{x}_l,t_0)e^{\imagI
\theta_l}e^{-\imagI\phi_L(\v{x}_l)}e^{\imagI(\v{p}_l-\v{k})\cdot(\v{x}-\v{x}_l)}
\right\}\\\nonumber &+& \frac{1}{\sqrt{2}}\ket{\InternalState_2}
\left\{c_2(\v{x}-\v{x}_l,t_0)e^{\imagI
\theta_l}e^{\imagI\v{p}_l\cdot(\v{x}-\v{x}_l)} - \imagI\,
c_1(\v{x}-\v{x}_u,t_0)e^{\imagI
\theta_u}e^{\imagI\phi_L(\v{x}_u)}e^{\imagI(\v{p}_u+\v{k})\cdot(\v{x}-\v{x}_u)}
\right\}
\end{eqnarray}
The probability of finding the atom in either output port
$\ket{\InternalState_1}$ or $\ket{\InternalState_2}$ is
\begin{eqnarray}
\abs{\bra{\InternalState_1}\braket{\v{x}}{\TotalState(t)}}^2 &=& \frac{\abs{c_1}^2+\abs{c_2}^2}{2}+\frac{1}{2}\left(\imagI\,c_1\,c_2^\ast\,e^{\imagI\Delta\phi_1}+c.c.\right)\\
\abs{\bra{\InternalState_2}\braket{\v{x}}{\TotalState(t)}}^2 &=&
\frac{\abs{c_1}^2+\abs{c_2}^2}{2}-\frac{1}{2}\left(\imagI\,c_1\,c_2^\ast\,e^{\imagI\Delta\phi_2}+c.c.\right)
\end{eqnarray}
with $c_1\equiv c_1(\v{x}-\v{x}_u,t_0)$ and $c_2\equiv
c_2(\v{x}-\v{x}_l,t_0)$. For the total phase shift we find
\begin{eqnarray}
\Delta\phi_1 &\equiv & \Big{\{}\theta_u+\v{p}_u\cdot(\v{x}-\v{x}_u)\Big{\}}-\Big{\{}\theta_l-\phi_L(\v{x}_l)+(\v{p}_l-\v{k})\cdot(\v{x}-\v{x}_l)\Big{\}}\label{DeltaPhi-1-raw}\\
&=&\underbrace{\theta_u-\Big{(}\theta_l-\phi_L(\v{x}_l)\Big{)}}_{\Delta\phi_\text{propagation,1}~+~\Delta\phi_\text{laser,1}}+\underbrace{\Big{.}\,\,\v{\bar{p}}_1\cdot\v{\Delta
x}\,\,}_{\Delta\phi_\text{separation,1}}+\,\,\v{\Delta
p}\cdot(\v{x}-\v{\bar{x}})\label{DeltaPhi-1}
\end{eqnarray}
and
\begin{eqnarray}
\Delta\phi_2 &\equiv & \Big{\{}\theta_u + \phi_L(\v{x}_u) +
(\v{p}_u+\v{k})\cdot(\v{x}-\v{x}_u) \Big{\}}-
\Big{\{} \theta_l + \v{p}_l\cdot(\v{x}-\v{x}_l) \Big{\}}\label{DeltaPhi-2-raw}\\
&=&\underbrace{\Big{(}\theta_u +
\phi_L(\v{x}_u)\Big{)}-\theta_l}_{\Delta\phi_\text{propagation,2}~+~\Delta\phi_\text{laser,2}}+\underbrace{\Big{.}\,\,\v{\bar{p}}_2\cdot\v{\Delta
x}\,\,}_{\Delta\phi_\text{separation,2}}+\,\,\v{\Delta
p}\cdot(\v{x}-\v{\bar{x}})\label{DeltaPhi-2}
\end{eqnarray}
where $\v{\bar{p}}_1=\frac{\v{p}_u + (\v{p}_l-\v{k})}{2}$ and
$\v{\bar{p}}_2=\frac{(\v{p}_u+\v{k})+\v{p}_l}{2}$ are the average
momenta in the \ket{\InternalState_1} (slow) and
\ket{\InternalState_2} (fast) output ports, respectively.  In
general, the separation phase is \be\Delta\phi_\text{separation}=
\v{\bar{p}}\cdot\v{\Delta x}\label{SeparationPhase}\ee which
depends on the separation $\v{\Delta x}\equiv \v{x}_l - \v{x}_u$
between the centers of the wavepackets from each arm as well as
the average canonical momentum $\v{\bar{p}}$ in the output port.
We point out that even though the definitions
(\ref{DeltaPhi-1-raw}) and (\ref{DeltaPhi-2-raw}) use the same
sign convention as our previous expressions for laser
(\ref{LaserPhase}) and propagation (\ref{PropagationPhase}) phase
in the sense of $(~)_\text{upper}-(~)_\text{lower}$, the
separation vector $\v{\Delta x}$ is defined as
$(\v{x})_\text{lower}-(\v{x})_\text{upper}$.

Notice that the phase shift expressions (\ref{DeltaPhi-1}) and
(\ref{DeltaPhi-2}) contain a position dependent piece $\v{\Delta
p}\cdot(\v{x}-\v{\bar{x}})$, where $\v{\bar{x}}\equiv\frac{\v{x}_u
+ \v{x}_l}{2}$ and $\v{\Delta p}=(\v{p}_u+\v{k})-\v{p}_l=\v{p}_u -
(\v{p}_l-\v{k})$, owing to the fact that the contributions from
each arm may have different momenta after the last beamsplitter.
Typically this momentum difference is very small, so the resulting
phase variation has a wavelength that is large compared to the
spatial extent of the wavefunction.  Furthermore, this effect
vanishes completely in the case of spatially averaged detection
over a symmetric wavefunction.

Finally, we show that the total phase shifts $\Delta\phi_1$ and
$\Delta\phi_2$ for the two output ports are actually equal, as
required by conservation of probability.  According to Eqs.
(\ref{DeltaPhi-1}) and (\ref{DeltaPhi-2}), the contributions to
the total phase differ in the following ways:
\begin{eqnarray}
\nonumber\Big{(}\Delta\phi_\text{propagation,1}+\Delta\phi_\text{laser,1}\Big{)}-\Big{(}\Delta\phi_\text{propagation,2}+\Delta\phi_\text{laser,2}\Big{)}&=&\phi_L(\v{x}_l)-\phi_L(\v{x}_u)\\
\nonumber &=& \v{k}\cdot(\v{x}_l-\v{x}_u)=\v{k}\cdot\v{\Delta x}
\end{eqnarray}
\be\nonumber\Delta\phi_\text{separation,1}-\Delta\phi_\text{separation,2}=\v{\bar{p}}_1\cdot\v{\Delta
x} - \v{\bar{p}}_2\cdot\v{\Delta x} = -\v{k}\cdot\v{\Delta x}\ee
Together these results imply that $\Delta\phi_1=\Delta\phi_2$ and
prove that the total interferometer phase shift $\Delta
\phi_\text{tot}$ is independent of the output port.

The accuracy of the above formalism is dependent on the
applicability of the aforementioned stationary phase approximation
as well as the short pulse limit.  The stationary phase
approximation breaks down when the external Hamiltonian varies
rapidly compared to the phase space width of the atom wavepacket.
The short pulse limit requires that the atom's velocity not change
appreciably during the duration of the atom-light interaction.
Both approximations are justified to a large degree for a typical
light pulse atom interferometer, but in the most extreme high
precision applications such as we consider here, important
corrections are present.  However, we emphasize that these errors
due to finite pulse duration and wavepacket size are well-known,
previously established backgrounds.

\section{Applications in inertial navigation}
The navigation problem is easily stated:  How do we determine a
platform's trajectory as a function of time?  In the 20th century,
solutions to the problem have led to the development of
exquisitely refined hardware systems and navigation algorithms.
Today we take for granted that a hand-held GPS reciever can be
used to obtain meter level position determination.  When GPS is
unavailable (for example, when satellites are not in view),
position determination becomes much less accurate.  In this case,
stand-alone ``black-box'' inertial navigation systems, comprised
of a suite of gyroscope and acclerometers, are used to infer
position changes by integrating the outputs of these inertial
force sensors. State-of-the-art commercial grade navigation
systems have position drift errors of kilometers per hour of
navigation time, significantly worse than the GPS solution.  How
can we close the performance gap between GPS and inertial systems?
The way forward is improved instrumentation: better gyroscopes and
accelerometers.

The light-pulse interferometry method is well suited to inertial
applications.  As shown in Section \ref{Sec: AI answer}, the phase
shift for the light pulse interferometer consists of contributions
from path phases, optical interactions, and separation phases.
However, for the sensitivity range of interest, the optical phase
shifts dominate, and therefore $\Delta \phi_{tot} \approx \Delta
\phi_{laser}$. In this case there is a straightforward
interpretation for the operation of the sensor:  the sensor
registers the time evolution of the relative position of the
atomic wavepackets with respect to the sensor case (defined by the
opto-mechanical hardware for the laser beams) using optical
interferometry. Since distances are measured in terms of the
wavelength of light, and the atom is in a benign environment
(spurious forces, such as from magnetic field gradients, are
extremely small -- below $10^{-10} g$), the sensors are
characterized by superbly stable, low-noise operation.

The phase shift between the two paths is inferred by measuring the
probability of finding the atoms in a given output port.  Since we
are only concerned with the optical phase shift, we can calculate
the sensor output using Eq. \eqref{LaserPhaseIntro}. We see that
for a standard $\pi/2 - \pi- \pi/2$ excitation sequence
\cite{Kasevich_1991}, $\Delta \phi_{laser} =  {\bf k}_1 \cdot
({\bf x}_1^u-{\bf x}_1^0) - {\bf k}_2 \cdot ({\bf x}_2^u-{\bf
x}_2^0) - {\bf k}_2 \cdot ({\bf x}_2^l-{\bf x}_2^0) + {\bf k}_3
\cdot ({\bf x}_3^l-{\bf x}_3^0)$ . Here the subscript indexes each
of the three successive optical interactions, ${\bf x}_i^u$ and
${\bf x}_i^l$ are the semi-classical positions of the atom along
the upper and lower paths, respectively, at the time of each
interaction in a non-rotating, inertial coordinate system, ${\bf
x}_i^0$ is the phase reference for the optical fields, and ${\bf
k}_i$ is the propagation vector of the laser field associated with
each pulse.

\subsection{Gyroscope}
Assuming the atoms have initial velocity ${\bf v}_0$ and the
effective Raman propagation vectors initially have common
orientations ${\bf k}_0$ which rotate with angular rate ${\bf
\Omega}$, it is straightforward to show that $\Delta \phi_{laser}
\approx {\bf k}_0 \cdot (2 {\bf v}_0 \times {\bf \Omega})T^2$.
This configuration is well-suited to precision measurements of
platform rotations.  This expression can be put in a form
analogous to the Sagnac shift for optical interferometers by
noting that for small rotation rates $\Omega$, $\v{A} =
\frac{\hbar}{m}({\bf k_0}\times {\bf v_0}) T^2$ is the area
enclosed by the interfering paths. Thus this shift can also be
written in the Sagnac form $\phi =
\frac{2m}{\hbar}\v{\Omega}\cdot\v{A}$ -- proportional to the
product of the enclosed area and rotation rate.

\begin{figure}
\begin{center}
\includegraphics[width=300pt]{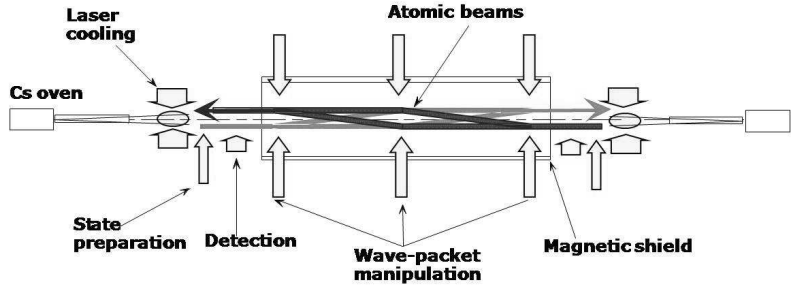}
\caption{ \label{Fig:gyroschematic} Schematic illustration of an
atomic beam gyroscope.}
\end{center}
\end{figure}

\begin{figure}
\begin{center}
\includegraphics[width=300pt]{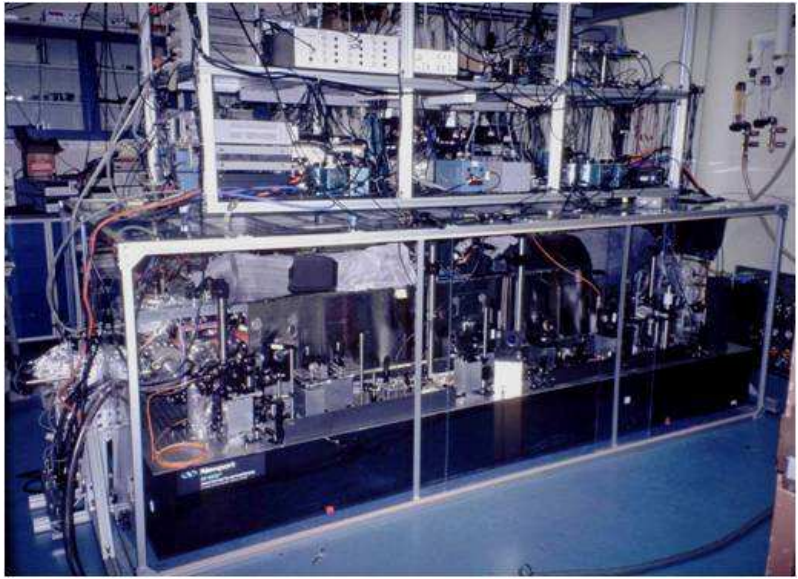}
\caption{ \label{Fig:gyrophoto} Photograph of gyroscope.}
\end{center}
\end{figure}

\begin{figure}
\begin{center}
\includegraphics[width=300pt]{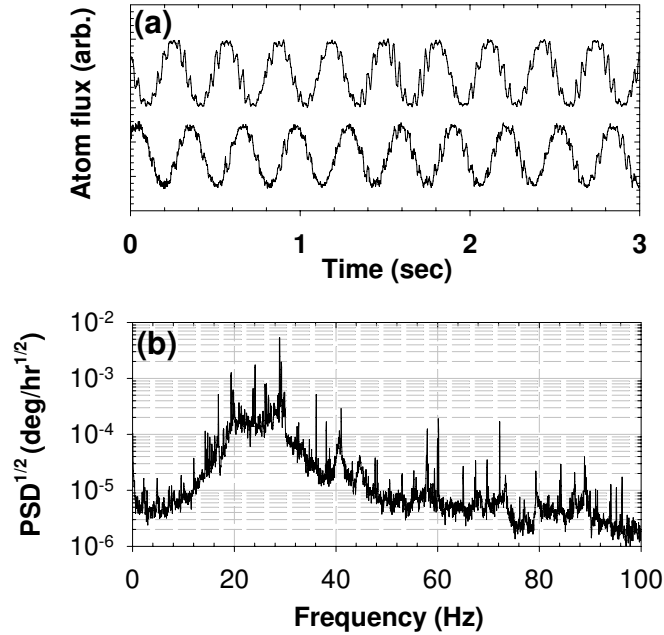}
\caption{ \label{Fig:gyronoise} Gyroscope interference fringes (a)
and power spectral density of gyroscope noise (b).  The increase
in noise in the 10-50 Hz band is due to technical rotation noise
sources in the building (the gyroscope was mounted directly to the
building floor).  See Ref. \cite{Durfee}.}
\end{center}
\end{figure}

\begin{figure}
\begin{center}
\includegraphics[width=300pt]{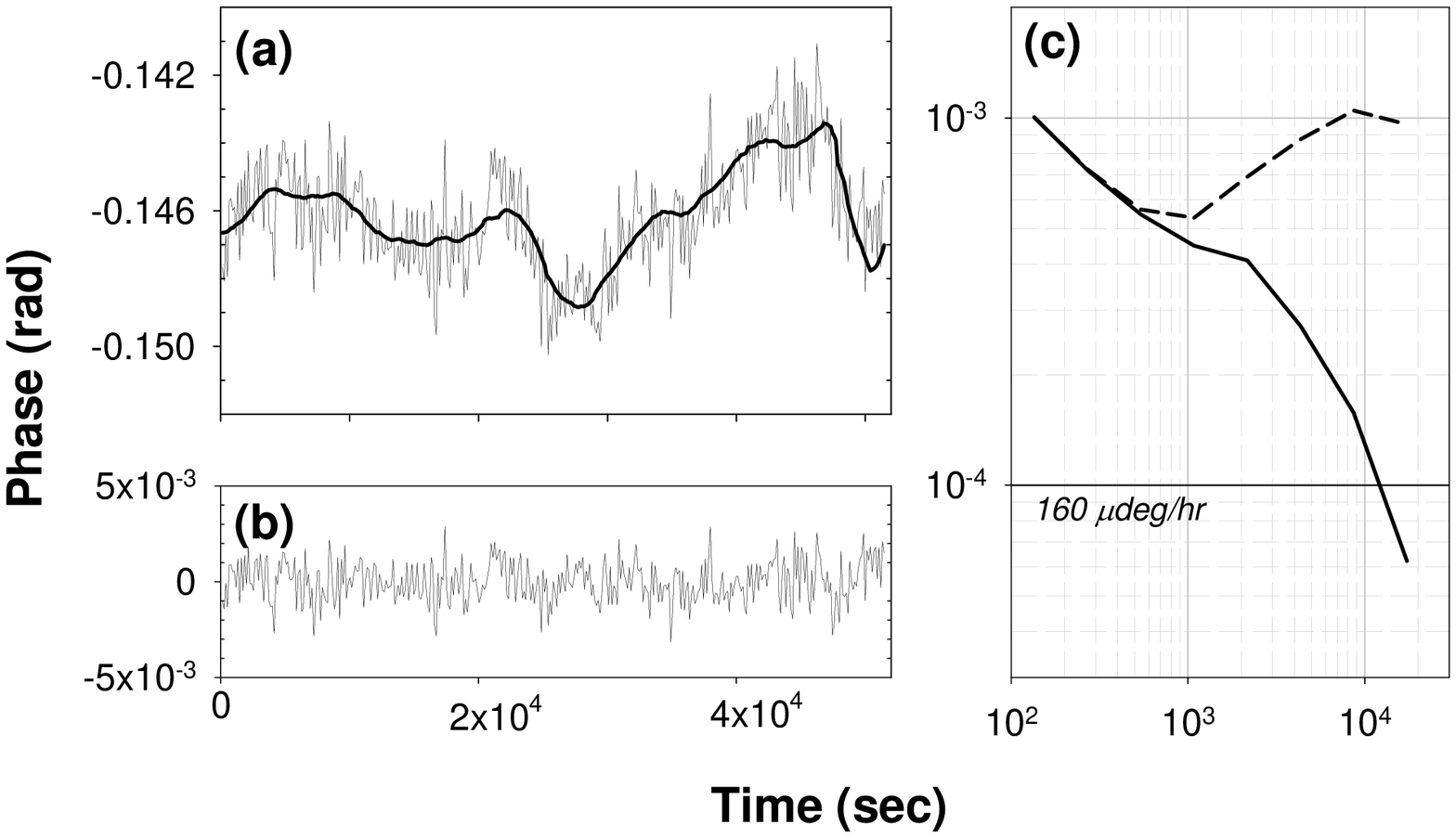}
\caption{ \label{Fig:gyrostability} (a) Drift in case reversed
gyroscope output with time and its correlation with instrument
temperature (solid black).  (b) Gyro noise output residual after
subtracting temperature model.  (c) Allan deviation of gyro output
before and after temperature compensation (see Ref.
\cite{Durfee}).}

\end{center}
\end{figure}

Gyroscopes built on this principle have achieved performance
levels in the laboratory which compare favorably with state-of-the
art gyroscopes, as shown in Figs. \ref{Fig:gyroschematic},
\ref{Fig:gyrophoto}, \ref{Fig:gyronoise} and
\ref{Fig:gyrostability}. Key figures of merit for gyroscope
performance include gyroscope noise (often referred to as angle
random walk), bias stability (stability of output for a null
input) and scale factor stability (stability of the multiplier
between the input rotation rate and output phase shift).  The
laboratory gyroscope illustrated below has achieved a demonstrated
angle random walk of $3~\mu\text{deg}/\text{hr}^{1/2}$, bias
stability of $60~\mu\text{deg}/\text{hr}$ (upper limit) and scale
factor stability of 5 ppm (upper limit).  Key drivers in the
stability of the gyroscope outputs are the stability of the
intensities used to drive the Raman transitions, and alignment
stability of the Raman beam optical paths.  Non-inertial phase
shifts associated with magnetic field inhomogeneities and spurious
AC Stark shifts are nulled using a case reversal technical where
the propagation directions of the Raman beams are periodically
reversed using electo-optic methods.

\subsection{Accelerometer}
\begin{figure}
\begin{center}
\includegraphics[width=200pt]{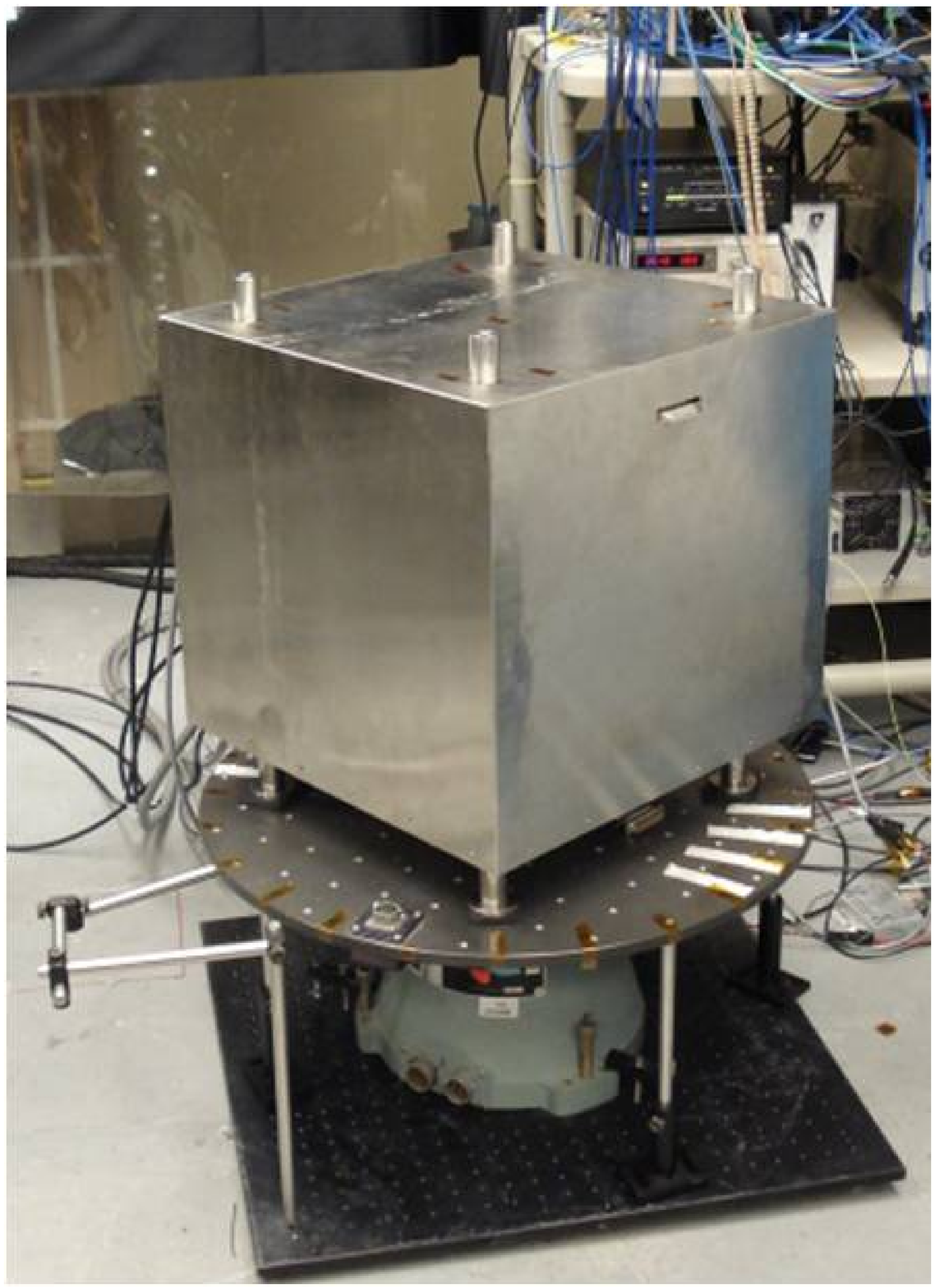}
\caption{ \label{Fig:compactaccel} Compact accelerometer.  This
instrument supports accelerometer, gyroscope and gravity
gradiometer operation modes.  The compact accelerometer has
demonstrated microGal sensitivity.}
\end{center}
\end{figure}

If the platform containing the laser beams accelerates, or if the
atoms are subject to a gravitational acceleration, the laser phase
shift then contains acceleration terms $\Delta \phi = {\bf k}_0
\cdot {\bf a} T^2.$  For a stationary interferometer, with the
laser beams vertically directed, this phase shift measures the
acceleration due to gravity. Remarkably, part per billion level
agreement has been demonstrated between the output of an atom
interferometer gravimeter and a conventional,
``falling-corner-cube'' gravimeter. \cite{AchimMetrologia}. In the
future, compact, geophysical ($10^{-8}$ $g$ accuracy) grade
instruments should enable low cost gravity field surveys (see Fig.
\ref{Fig:compactaccel}). For this type of instrument, laser
cooling methods are used to initially prepare ensembles of roughly
$10^7$ atoms at kinetic temperatures approaching 1 $\mu$K. At
these low temperatures, the rms velocity spread of the atomic
ensembles is a few cm/sec. Cold atom ensembles are then launched
on ballistic trajectories. In this configuration, the time between
laser pulses exceeds 100 msec, which means wavepackets separate by
roughly 1 mm over the course of the interferometer sequence.  The
phase shift is read-out by detecting the number of atoms in each
final state using resonance fluorescence and normalized detection
methods.

In general, both rotation and acceleration terms are present in
the sensor outputs.  For navigation applications, the rotation
response needs to be isolated from the acceleration response.  In
practice, this is accomplished by using multiple atom sources and
laser beam propagation axes.  For example, for the gyroscope
illustrated in Fig. \ref{Fig:gyroschematic}, counter-propagating
atom beams are used to isolate rotation induced phase shifts from
acceleration induced shifts.  It is interesting to note that the
same apparatus is capable of simultaneous rotation and
acceleration outputs -- a significant benefit for navigation
applications which require simultaneous output of rotation rate
and acceleration for three mutually orthogonal axes.  Since
gyroscope and accelerometer operation rest on common principles
and common hardware implementations, integration of sensors into a
full inertial base is straightforward.  Of course, particular
hardware implementations depend on the navigation platform (e.g.
ship, plane, land vehicle) and trajectory dynamics.

\subsection{Gravity gradiometer}
There is an additional complication in navigation system
architecture for high accuracy navigation applications: the
so-called ``problem of the vertical.''  Terrestrial navigation
requires determining platform position in the gravity field of the
Earth. Due to the Equivalence Principle, navigation system
accelerometers do not distinguish between the acceleration due to
gravity and platform acceleration.  So in order to determine
platform trajectory in an Earth-fixed coordinate system, the local
acceleration due to gravity needs to be subtracted from
accelerometer output in order to determine the acceleration of the
vehicle with respect to the Earth.  This means that the local
acceleration due to gravity needs to be independently known.  For
example, existing navigation systems use a gravity map to make
this compensation.  However, in present systems, this map does not
have enough resolution or accuracy for meter-level position
determination.  To give a feeling for orders of magnitude, a
$10^{-7}$ error in knowledge of the local acceleration due to
gravity integrates to meter-level position errors in 1 hour.

There are at least two paths forward: 1) better maps or 2)
on-the-fly gravity field determination.  Improved maps can be
obtained with more precise surveys.  On-the-fly determination
seems impossible, due to the Equivalence Principle (since platform
accelerations cannot be discriminated from the acceleration due to
gravity). However, the outputs from a gravity gradiometer -- an
instrument which measures changes in the acceleration due to
gravity over fixed baselines -- can be used for this purpose.  The
idea is to integrate the gravity gradient over the inferred
trajectory to determine gravity as a function of position.  In
principle, such an instrument can function on a moving platform,
since platform accelerations cancel as a common mode when the
output from spatially separated accelerometers are differenced to
obtain the gradient.  In practice, such a strategy places hard
requirements on the stability of the component accelerometers:
their responses need to be matched to an exceptional degree in
order to discriminate gravity gradient induced accelerations
(typically below 10$^{-9}$ $g$) from other sensor error sources.

\begin{figure}
\begin{center}
\includegraphics[width=150pt]{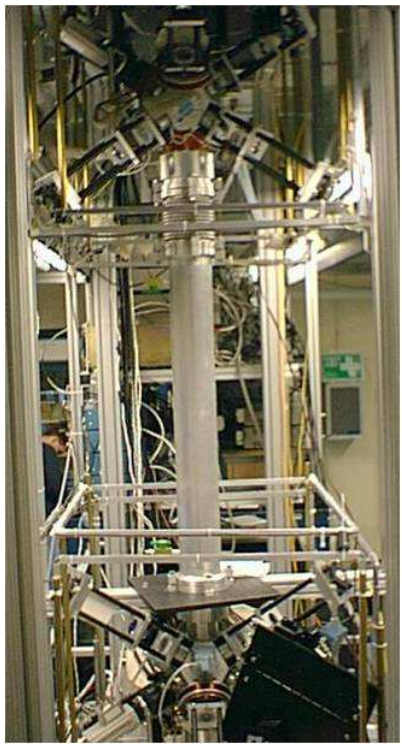}
\caption{ \label{Fig:ggphoto} Vertical axis gravity gradiometer.
Two atom interferometer accelerometers separated along a vertical
measurement axis are simultaneously interrogated by a common set
of Raman laser beams.  This apparatus was used to measure $G$,
Newton's constant \cite{Fixler}.}
\end{center}
\end{figure}

Due to the stability of their acceleration outputs, a pair of
light pulse accelerometers is well-suited to gravity gradient
instrumentation.  The basic idea is to simultaneously create two
spatially separated interferometers using a common laser beam. In
this way, technical acceleration noise of the measurement platform
is a common-mode noise source which leads to near identical phase
shifts in each accelerometer.  On the other hand, a gravity
gradient along the measurement axis results in a residual
differential phase shift.  This configuration has been used to
measure the gravity gradient of the Earth, as well as the gravity
gradient associated with nearby mass distributions, as illustrated
in Fig. \ref{Fig:ggphoto}.  Laboratory gravity gradiometers have
achieved resolutions below 1 E (where 1 E = $10^{-9}$ sec$^{-2}$).
This configuration has also been used to measure the Newtonian
constant of gravity $G$ \cite{Fixler,FixlerThesis,Tino2006}. In
future navigation systems, an ensemble of accelerometers,
configured along independent measurement axes, could acquire the
full gravity gradient tensor.

\begin{figure}
\begin{center}
\includegraphics[width=200pt]{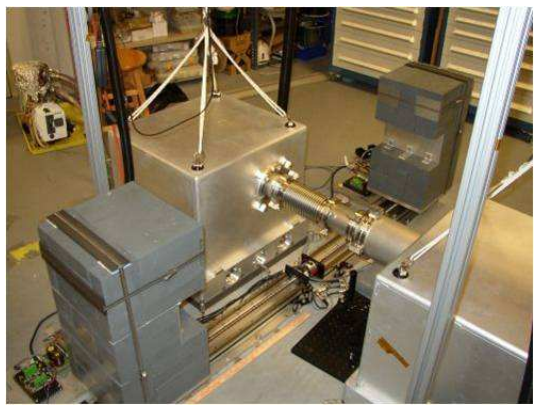}
\caption{ \label{Fig:newggphoto} Gravity gradiometer with
horizontal measurement axis.  Raman beams propagate along the
axis defined by the tube connecting each accelerometer housing.}
\end{center}
\end{figure}

\begin{figure}
\begin{center}
\includegraphics[width=350pt]{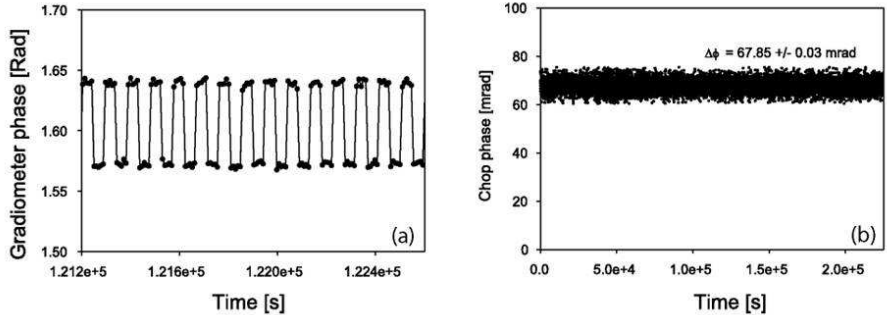}
\caption{ \label{Fig:ggdata} Gravity gradiometer response to a
proof mass which is periodically brought in close proximity to one
of the accelerometer regions.  For the proof masses used for this
demonstration, the apparatus is capable of resolving changes in
$G$ at the $3\times 10^{-4}$ level \cite{Biedermann}.}
\end{center}
\end{figure}

In addition to their role as navigation aids, gravity gradiometers
have applications in geodesy and oil/mineral exploration.  The
idea here is that mass/density anomalies associated with
interesting geophysical features (such as kimberlite pipes -- in
the case of diamond exploration -- or salt domes -- in the case of
oil exploration) manifest as gravity anomalies.  In some cases,
these anomalies can be pronounced enough to be detected by a
gravity gradiometer from an airborne platform.  Atom-based gravity
gradiometers appear to have competitive performance figures of
merit for these applications as compared with existing
technologies.  For these applications, the central design
challenge is realization of an instrument which has very good
noise performance, but also is capable of sustained operation on a
moving platform.  Figs. \ref{Fig:newggphoto}, \ref{Fig:ggdata}
illustrate a system currently under development for this purpose.

\section{Application to tests of the Equivalence Principle}
Precision tests of the Equivalence Principle (EP) promise to
provide insight into fundamental physics.  Since the EP is one of
the central axioms of general relativity (GR), these experiments
are powerful checks of gravity and can tightly constrain new
theories.  Furthermore, EP experiments test for hypothetical fifth
forces since many examples of new forces are
EP--violating\cite{fifthForce}.

The Equivalence Principle has several forms, with varying degrees
of universality.  Here we consider tests of the Weak Equivalence
Principle, which can be stated as follows: the motion of a body in
a gravitational field in any local region of space-time is
indistinguishable from its motion in a uniformly accelerated
frame.  This implies that the body's inertial mass is equal to its
gravitational mass, and that all bodies locally fall at the same
rate under gravity, independent of their mass or composition.

The results of EP experiments are typically expressed in terms of
the E\"{o}tv\"{o}s parameter $\eta=\Delta a/\bar{a}$, where
$\Delta a$ is the EP violating differential acceleration between
the two test bodies and $\bar{a}$ is their average acceleration
\cite{cliffwillLR}.  Currently, two conceptually different
experiments set the best limits on the Equivalence Principle.
Lunar Laser Ranging (LLR), which tests the EP by comparing the
acceleration of the Earth and Moon as they fall toward the Sun,
limits EP violation at $\eta=(-1.0 \pm 1.4)\times 10^{-13}$
\cite{LLRwilliams}. Recently, the E\"{o}t-Wash group has set a
limit of $\eta=(0.3 \pm 1.8)\times 10^{-13}$ using an Earth-based
torsion pendulum apparatus\cite{AdelbergerEP}.  Several proposed
satellite missions aim to improve on these limits by observing the
motion of macroscopic test bodies in orbit around the
Earth\cite{step, microscope}. Here we discuss our effort to
perform a ground-based EP test using individual atoms with a goal
of measuring $\eta\sim 10^{-15}$.   Instead of macroscopic test
masses, we compare the simultaneous acceleration under gravity of
freely-falling cold atom clouds of $^{87}$\!Rb and $^{85}$\!Rb
using light-pulse atom interferometry \cite{KasevichChu}.

Light-pulse atom interferometers have already been used to make
extremely accurate inertial force measurements in a variety of
configurations, including gyroscopes, gradiometers, and
gravimeters.  For example, the local gravitational acceleration
$g$ of freely-falling Cs atoms was measured with an accuracy $\sim
10^{-9}g$ \cite{AchimMetrologia}.  Gravity gradiometers have been
used to suppress noise as well as many systematic errors that are
present in absolute $g$ measurements by comparing the acceleration
of two displaced samples of atoms.  A differential measurement of
this kind was used to measure the Newtonian constant of gravity
$G$ with an accuracy of $\sim 3\times 10^{-3}G$ \cite{Fixler}. The
EP measurement we describe here benefits from an analogous
differential measurements strategy, where in this case the
common-mode noise suppression arises from a comparison between two
co-located isotopes of different mass, rather than between
spatially separated atoms as in a traditional gradiometer.

\subsection{Proposed experiment overview}
\label{Sec: EP experiment overview}

\begin{figure}
\begin{center}
\includegraphics[width=350pt]{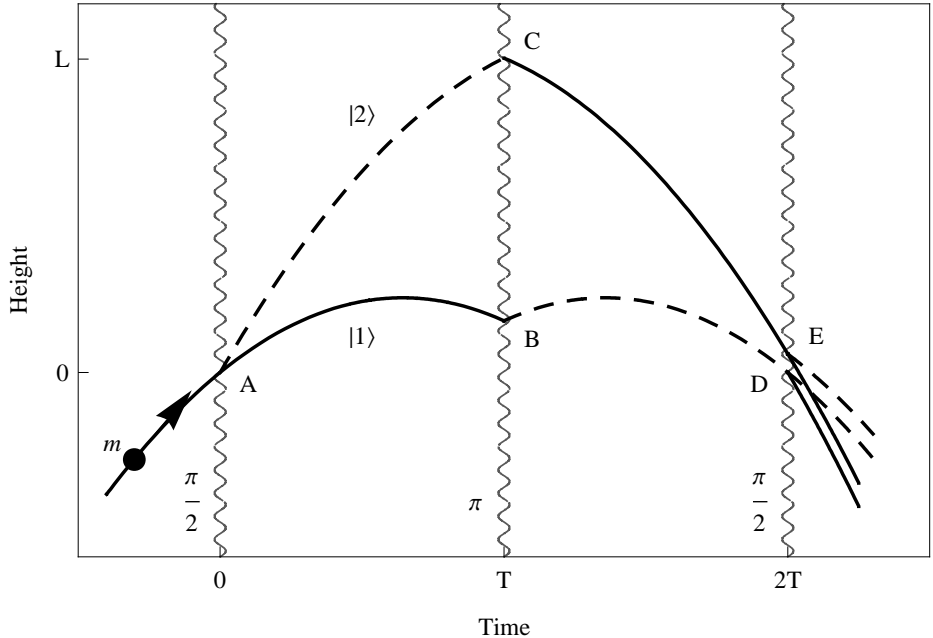}
\caption{\label{Fig:SpaceTime} Space--time diagram for a single
atom of mass $m$ during the interferometer pulse sequence. The
atom is launched with velocity $v_z$ from the bottom of the vacuum
system.  At time $t=0$, a $\frac{\pi}{2}$ (beamsplitter) pulse is
applied to coherently divide the atom wavefunction. After a time
$T$, a $\pi$ (mirror) pulse is applied that reverses the relative
velocity between the wavefunction components.  A final
$\frac{\pi}{2}$ (beamsplitter) pulse at time $2T$ results in
interference between the two space--time paths.  The
interferometer phase shift is inferred by measuring the
probability of detecting the atom in either state $\ket{1}$ (solid
line) or state $\ket{2}$ (dashed line).  Note that points $D$ and
$E$ are in general spatially separated in the presence of
non-uniform forces, leading to a separation phase shift.}
\end{center}
\end{figure}

The proposed experiment ideally consists of simultaneously
observing the free-fall motion of the two Rb isotopes in the
absence of all non-gravitational forces.  To this end, the
measurement is performed inside a $10~\text{cm}$ diameter by
$8.8~\text{m}$ long cylindrical ultra high vacuum chamber.  To
maximize their free-fall time, the atoms are launched in a
vertical fountain geometry from the bottom of the chamber.
Light-pulse atom interferometry is performed while the atoms are
in free-fall, and the resulting phase shift is sensitive to the
atoms' acceleration. Figure \ref{Fig:SpaceTime} is a space-time
diagram depicting the trajectories that each atom follows during
the free-fall interferometry sequence.

To maximize the cancellation of spurious effects, both the
$^{87}$\!Rb and $^{85}$\!Rb atom clouds are launched at the same
time and are made to follow the same trajectories as closely as
possible.  We launch both isotopes from the same magnetic trap in
order to minimize any differences between their initial positions
and velocities.  As a result of the small isotope shift between
$^{87}$\!Rb and $^{85}$\!Rb, we are able to use the same laser
pulses to simultaneously manipulate them during the interferometer
sequence (see Fig. \ref{Fig:SpaceTime}).  Using the same laser
makes the apparatus insensitive to pulse timing jitter and
dramatically reduces the phase noise stability requirements of the
lasers.

A single measurement of acceleration in our atom interferometer
consists of three steps: atom cloud preparation, interferometer
pulse sequence, and detection.  In the first step a
sub-microkelvin cloud of $\sim 10^7$ atoms is formed using laser
cooling and evaporative cooling in a TOP trap \cite{TOPtrap}. This
dilute ensemble of cold atoms is then launched vertically with
velocity $v_z\sim 13~\text{m}/\text{s}$ by transferring momentum
from laser light using an accelerated optical lattice potential
\cite{Phillips2002:JPhysB}.  This technique allows for precise
control of the launch velocity and, because it is a coherent
process, it avoids heating the cloud via spontaneous emission.

In the second phase of the measurement, the atoms follow free-fall
trajectories and the interferometry is performed.  A sequence of
laser pulses serve as beamsplitters and mirrors that coherently
divide each atom's wavepacket and then later recombine it to
produce interference (see Fig. \ref{Fig:SpaceTime}). The atom
beamsplitter is typically implemented using a stimulated
two-photon process (Raman or Bragg transitions), resulting in a
net momentum transfer of $\v{k_{\eff}}=\v{k_2}-\v{k_1}\approx
2\v{k_2}$ at each interaction. Since the acceleration sensitivity
of the interferometer is proportional to the effective momentum
$\hbar \keff$ transferred to the atom during interactions with the
laser, we intend to take advantage of more sophisticated atom
optics.  Large momentum transfer (LMT) beamsplitters with $\hbar
\keff=24\hbar k$ have been demonstrated \cite{HolgerLMT}, and up
to $100\hbar k$ may be possible.  Promising LMT beamsplitter
candidates include optical lattice manipulations
\cite{Phillips2002:JPhysB}, sequences of Raman pulses
\cite{McGuirk} and adiabatic passage methods \cite{Chu}.

The third and final step of each acceleration measurement is atom
detection.  At the end of the interferometer sequence, each atom
is in a superposition of the two output velocity states, as shown
by the diverging paths on the right in Fig. \ref{Fig:SpaceTime}.
These two final velocity states are directly analogous to the two
output ports of a Mach-Zehnder light interferometer after the
final recombining beamsplitter.  As with a light interferometer,
the probability that an atom will be found in a particular output
port depends on the relative phase acquired along the two paths of
the atom interferometer.  Since the output states differ in
velocity by $\sim \hbar\keff/m$, they spatially separate over
time.  After an appropriate drift time, the two velocity groups
can be separately resolved, and the populations can be then
measured by fluorescence imaging.

We now consider the expected sensitivity of our differential
$^{87}$\!Rb--$^{85}$\!Rb accelerometer.  Recent atom
interferometers have demonstrated sensor noise levels limited only
by the quantum projection noise of the atoms (atom shot noise)
\cite{clocks}.  Assuming a time--average atom flux of
$n=10^{6}~\text{atoms}/\text{s}$, the resulting shot
noise--limited phase sensitivity is $\sim
\frac{1}{\sqrt{n}}=10^{-3}~\text{rad}/\sqrt{\text{Hz}}$. The phase
shift in an atom interferometer due to a constant acceleration $g$
is $\Delta\phi=\keff g T^2$ \cite{bongs}.  Taking advantage of the
$L\approx 8.8~\text{m}$ vacuum system allows for a long
interrogation time of up to $T=1.34~\text{s}$.  Finally, using
$\hbar \keff=10\hbar k$ LMT beamsplitters results in an
acceleration sensitivity of $\sim 7\times
10^{-13}~g/\sqrt{\text{Hz}}$ and a precision of $<10^{-15}g$ after
$\sim 1~\text{day}$ of integration.  In the most conservative
case, constraining ourselves to conventional $2\hbar k$ atom
optics leads to a precision of $<10^{-15}g$ after $\sim
1~\text{month}$ of integration. This estimate is based on
realistic extrapolations from current performance levels, which
are at $10^{-10}g$ \cite{Fixler}.

\subsection{Error model}
\label{Sec:EP phase shift calc}

An accurate test of the EP requires a thorough understanding of
potential backgrounds.  To reach the goal sensitivity, we must
control spurious accelerations to $< 10^{-15}g$.  Systematic
errors at this level can arise from many sources, including
gravity gradients, Earth's rotation, and electromagnetic forces.
To calculate these contributions to the phase shift, we follow the
prescription outlined in Section \ref{Sec:AI Calc}.  We take the
atom's Lagrangian in the lab frame to be \be L=\frac{1}{2}m
(\dot{\v{r}}+\v{\Omega}\times(\v{r}+\v{R}_e))^2-m
\phi(\v{r}+\v{R}_e)-\frac{1}{2}\alpha
\v{B(\v{r})}^2\label{Eq:Lagrangian}\ee where $\v{r}$ is the
position of the atom in the lab frame, $\v{R}_e=(0, 0, R_e)$ is
the radius of the Earth, $\v{\Omega}=(0, \Omega_y, \Omega_z)$ is
the Earth's rotation rate, and $\phi(\v{r})$ is the gravitational
potential. In the chosen coordinate system, $z$ is the vertical
direction in the lab and $\v{\Omega}$ lies in the $y$-$z$ plane.
We then expand $\phi$ in a Taylor series about $\v{R}_e$, \be
\phi(\v{r}+\v{R}_e) = -\left(\v{g}\cdot\v{r} + \frac{1}{2!}
(T_{ij})r_i r_j + \frac{1}{3!} (Q_{ijk})r_i r_j r_k + \frac{1}{4!}
(S_{ijkl})r_i r_j r_k r_l\right)
\label{Eq:GravitationalTaylorSeries} \ee where Earth's gravity
field is $\v{g}\equiv - \nabla\phi(\v{R}_e)$, the gravity gradient
tensor is $T_{ij}\equiv \partial_j g_{i}$, the second gradient
tensor is $Q_{ijk}\equiv \partial_{k}\partial_{j} g_{i}$, the
third gradient tensor is $S_{ijkl}\equiv
\partial_{l}\partial_{k}\partial_{j} g_{i}$, and repeated indices are summed over.
Since $\v{\hat{z}}$ is in the vertical direction in the lab we
have that $\v{g}=(0,0,-g)$ and $g=9.8~\text{m}/\text{s}^2$. The
interferometer follows a fountain geometry which is approximately
one-dimensional along the $z$-direction, so we only include
$Q_{zzz}$ and $S_{zzzz}$ and safely ignore the other second and
third gradient tensor terms.  Likewise, in this analysis we assume
that off-diagonal gradient tensor terms $T_{ij}$ with $i\neq j$
are small and can be ignored (this is exactly true for a perfectly
spherical Earth).  The effects of higher-order moments of the
gravitational field are treated separately using a perturbative
calculation as described in Section \ref{Sec:Gravity
Inhomogeneities}.

Because magnetic fields can cause significant systematic errors,
the atoms are prepared in one of the magnetic field insensitive
clock states (\ket{m_F=0} states).  The residual energy shift in a
magnetic field $\v{B}$ is then $U_B=\frac{1}{2}\alpha\v{B}^2$,
where $\alpha$ is the second order Zeeman shift coefficient. We
consider magnetic fields of the form \be \v{B(\v{r})} = \left(B_0
+\frac{\partial B}{\partial z} z\right)\v{\hat{z}}
\label{Eq:TaylorMagneticFields}\ee where $B_0$ is a constant bias
magnetic field, and $\frac{\partial B}{\partial z}$ is the
gradient of the background magnetic field.  While this linear
model is sufficient for slowly varying fields, in Section
\ref{Sec:Magnetic Inhomgeneities} we describe a perturbative
calculation that can account for more complicated magnetic field
spatial profiles.

We do not include additional electromagnetic forces in the
Lagrangian as their accelerations are well below our systematic
threshold.  For neutral atoms, electric fields are generally not a
concern since the atom's response is second order. Furthermore,
electric fields are easily screened by the metallic vacuum
chamber, leading to negligibly small phase shifts.  Short range
effects due to the Casimir \cite{casimir} force or local patch
potentials \cite{patch} are also negligible since the atoms are
kept far ($>1~\text{cm}$) from all surfaces throughout the
experiment.

As explained in Section \ref{Sec:AI proof}, we point out that the
phase shift derived from Eq. \ref{Eq:DeltaPhiTotal} is only
exactly correct for Lagrangians that are second order in position
and velocity.  When this is not true, as is the case in Eq.
\ref{Eq:Lagrangian} when $Q_{ijk}\neq 0$ and $S_{ijkl}\neq 0$, the
semiclassical formalism breaks down and there are quantum
corrections to the phase shift.  However, these corrections depend
on the size of the atom wavepacket compared to the length scale of
variation of the potential, and are typically negligible for
wavepackets $\sim 1~\text{mm}$ in size.

To analytically determine the trajectories $\v{r}(t)$, we solve
the Euler-Lagrange equations using a power series expansion in
$t$: \be r_i(t) = \sum_{n=0}^N a_{in} (t-t_0)^n\qquad\qquad
(i=1,2,3)\ee The coefficients $a_{in}$ are determined recursively
after substitution into the equations of motion.  This expansion
converges quickly as long as $\Omega T \ll 1$ and
$\left|\frac{r^n}{g}\frac{\partial^n g}{\partial r^n}\right|\ll
1$.  For our apparatus with characteristic length $r\sim
10~\text{m}$ and time $T\sim 1~\text{s}$ these conditions are
easily met, since $\Omega T \sim 10^{-4}~\text{rad}$ and
$\left|\frac{r^n}{g}\frac{\partial^n g}{\partial
r^n}\right|\sim\left(\frac{r}{R_e}\right)^n\lesssim 10^{-6}$
assuming a spherical Earth.  With these trajectories and the
interferometer geometry shown in Fig. \ref{Fig:SpaceTime} we
obtain the following expressions for the phase shift in the slow
(state $\ket{1}$) output port:
\begin{eqnarray}
  \Delta \phi_\text{propagation} &=& \frac{1}{\hbar}((S_{AC}+S_{CE})-(S_{AB}+S_{BD})) \label{Eq:PropagationPhase}\\
  \Delta \phi_\text{laser} &=& \phi_L(\v{r}_A,0)-\phi_L(\v{r}_C,T)-\phi_L(\v{r}_B,T)+\phi_L(\v{r}_D,2T) \label{Eq:LaserPhase} \\
  \Delta \phi_\text{separation} &=& \frac{1}{2\hbar}(\v{p}_D + \v{p}_E)\cdot(\v{r}_D-\v{r}_E) \label{Eq:SeparationPhase}
\end{eqnarray}
where $S_{ij}$ is the classical action along the path segment
between points $\v{r}_i$ and $\v{r}_j$, and
$\v{p}_i=\partial_\v{\dot{r}}L(\v{r}_i)$ is the classical
canonical momentum at point $\v{r}_i$ after the final
beamsplitter.  The laser phase shift at each interaction point is
\be \phi_L(\v{r},t) = \v \keff \cdot \v r - \omega_\eff t + \phi_0
\ee where $\v\keff$ and $\omega_\eff$ are the effective
propagation vector and frequency, respectively, for whatever
atom--laser interaction is used to implement the atom optics.  In
the case of the stimulated two--photon processes mentioned
earlier, $\v{k_{\eff}}=\v{k_2}-\v{k_1}$ and
$\omega_\eff=\omega_2-\omega_1=(k_2-k_1)/c.$

\begin{table}
\begin{center}
\begin{math}
\begin{array}{lccc}
\hline\\
\qquad\qquad & \qquad\qquad\text{Phase shift}\qquad\qquad & \qquad\text{Size (rad)}\qquad & \qquad\text{Fractional size}\qquad \\
\hline\hline\\
 1 & -k_{\text{eff}} g T^2 & -2.85\times 10^8 & 1.00 \\
 2 & k_{\text{eff}} R_e \Omega_y^2 T^2 & 6.18\times 10^5 & 2.17\times 10^{\text{-3}}
   \\
 3 & -k_{\text{eff}} T_{zz} v_z T^3 & 1.58\times 10^3 & 5.54\times 10^{\text{-6}} \\
 4 & \frac{7}{12} k_{\text{eff}} g T_{zz} T^4  & -9.21\times 10^2 & 3.23\times 10^{\text{-6}} \\
 5 & -3 k_{\text{eff}} v_z \Omega_y^2 T^3  & -5.14 & 1.80\times 10^{\text{-8}} \\
 6 & 2 k_{\text{eff}}  v_x \Omega_y T^2 & 3.35 & 1.18\times 10^{\text{-8}} \\
 7 & \frac{7}{4} k_{\text{eff}} g \Omega_y^2 T^4  & 3.00 & 1.05\times 10^{\text{-8}} \\
 8 & -\frac{7}{12} k_{\text{eff}} R_e T_{zz} \Omega_y^2 T^4 & 2.00 & 7.01\times
   10^{\text{-9}} \\
 9 & -\frac{\hbar k_{\text{eff}}^2}{2 m}T_{zz} T^3   & 7.05\times 10^{\text{-1}} & 2.48\times
   10^{\text{-9}} \\
 10 & \frac{3}{4} k_{\text{eff}} g Q_{zzz} v_z T^5 & 9.84\times 10^{\text{-3}} &
   3.46\times 10^{\text{-11}} \\
 11 & -\frac{7}{12} k_{\text{eff}} Q_{zzz} v_z^2 T^4 & -7.66\times 10^{\text{-3}} &
   2.69\times 10^{\text{-11}} \\
 12 & -\frac{7}{4} k_{\text{eff}} R_e \Omega_y^4 T^4 & -6.50\times 10^{\text{-3}} &
   2.28\times 10^{\text{-11}} \\
 13 & -\frac{7}{4} k_{\text{eff}} R_e \Omega_y^2 \Omega_z^2 T^4 &
   -3.81\times 10^{\text{-3}} & 1.34\times 10^{\text{-11}} \\
 14 & -\frac{31}{120} k_{\text{eff}} g^2 Q_{zzz} T^6  & -3.39\times 10^{\text{-3}} & 1.19\times
   10^{\text{-11}} \\
 15 & -\frac{3\hbar  k_{\text{eff}}^2}{2 m}\Omega_y^2 T^3 & -2.30\times 10^{\text{-3}} &
   8.06\times 10^{\text{-12}} \\
 16 & \frac{1}{4} k_{\text{eff}} T_{zz}^2 v_z T^5 & 2.19\times 10^{\text{-3}} &
   7.68\times 10^{\text{-12}} \\
 17 & -\frac{31}{360} k_{\text{eff}} g T_{zz}^2 T^6 & -7.53\times 10^{\text{-4}} & 2.65\times
   10^{\text{-12}} \\
 18 & 3 k_{\text{eff}} v_y \Omega_y \Omega_z T^3 & 2.98\times
   10^{\text{-4}} & 1.05\times 10^{\text{-12}} \\
 19 & -k_{\text{eff}} \Omega_y \Omega_z y_0 T^2 & -7.41\times 10^{\text{-5}} &
   2.60\times 10^{\text{-13}} \\
 20 & -\frac{3}{4} k_{\text{eff}} R_e Q_{zzz} v_z \Omega_y^2 T^5 &
   -2.14\times 10^{\text{-5}} & 7.50\times 10^{\text{-14}} \\
 21 & \frac{31}{60} k_{\text{eff}} g R_e Q_{zzz} \Omega_y^2 T^6  & 1.47\times
   10^{\text{-5}} & 5.17\times 10^{\text{-14}} \\
 22 & \frac{3}{2} k_{\text{eff}} T_{zz} v_z \Omega_y^2 T^5 & -1.42\times
   10^{\text{-5}} & 5.00\times 10^{\text{-14}} \\
 23 & -\frac{7}{6} k_{\text{eff}} T_{zz} v_x \Omega_y T^4 & 1.08\times
   10^{\text{-5}} & 3.81\times 10^{\text{-14}} \\
 24 & -2 k_{\text{eff}} T_{xx} \Omega_y x_0 T^3 & -6.92\times 10^{\text{-6}} &
   2.43\times 10^{\text{-14}} \\
 25 & -\frac{7\hbar k_{\text{eff}}^2}{12 m} Q_{zzz} v_z T^4 & -6.84\times 10^{\text{-6}} &
   2.40\times 10^{\text{-14}} \\
 26 & -\frac{7}{6} k_{\text{eff}} T_{xx} v_x \Omega_y T^4 & -5.42\times
   10^{\text{-6}} & 1.90\times 10^{\text{-14}} \\
 27 & -\frac{31}{60} k_{\text{eff}} g T_{zz} \Omega_y^2 T^6 & 4.90\times 10^{\text{-6}}
   & 1.72\times 10^{\text{-14}} \\
 28 & k_{\text{eff}} T_{xx} v_z \Omega_y^2 T^5 & 4.75\times 10^{\text{-6}} &
   1.67\times 10^{\text{-14}} \\
 29 & \frac{3 \hbar k_{\text{eff}}^2}{8 m} g Q_{zzz} T^5 & 4.40\times 10^{\text{-6}} & 1.55\times
   10^{\text{-14}} \\
 30 & \frac{31}{360} k_{\text{eff}} R_e T_{zz}^2 \Omega_y^2 T^6 & 1.63\times
   10^{\text{-6}} & 5.74\times 10^{\text{-15}} \\
 31 & -\frac{31}{90} k_{\text{eff}} g T_{xx} \Omega_y^2 T^6 & -1.63\times
   10^{\text{-6}} & 5.74\times 10^{\text{-15}} \\
 32 & \frac{\hbar k_{\text{eff}}^2}{8 m} T_{zz}^2 T^5 & 9.78\times 10^{\text{-7}} & 3.43\times
   10^{\text{-15}} \\
 33 & -\frac{\hbar k_{\text{eff}} \alpha B_0  (\partial_z B) T^2}{m} & -7.67\times
   10^{\text{-8}} & 2.69\times 10^{\text{-16}} \\
 34 & \frac{31}{60} k_{\text{eff}} g S_{zzzz} v_z^2 T^6 & -7.52\times 10^{\text{-8}} &
   2.64\times 10^{\text{-16}} \\
 35 & -\frac{1}{4} k_{\text{eff}} S_{zzzz} v_z^3 T^5 & 3.64\times 10^{\text{-8}} &
   1.28\times 10^{\text{-16}} \\
 36 & \frac{31}{72} k_{\text{eff}} T_{zz} Q_{zzz} v_z^2 T^6 & -3.13\times
   10^{\text{-8}} & 1.10\times 10^{\text{-16}}\\
\hline
\end{array}
\end{math}
\caption{\label{Tab:full phases} Phase shift response for a single
atom interferometer $\frac{\pi}{2}-\pi-\frac{\pi}{2}$ sequence
given the Lagrangian in Eq. \ref{Eq:Lagrangian}.  Column 3 shows
the fractional size of each term compared to the acceleration
signal $k_{\text{eff}} g T^2$.  All terms with fractional phase
shift $> 10^{-16}$ are included.  The numbers are for a
$^{87}$\!Rb interferometer with the following parameters: $\keff =
2 k = 2\cdot\frac{2\pi}{780~\text{nm}}$, $T_{zz} = -2g/R_e$,
$T_{xx} = T_{yy} = g/R_e$, $Q_{zzz} = 6 g/R_e^2$, $S_{zzzz} = -24
g/R_e^3$, $R_e = 6.72 \times 10^6 \text{m}$, $B_0 =
100~\text{nT}$, and $\partial_z B = 0.1~\text{nT}/\text{m}$.  The
Earth's rotation rate is given by $\Omega_y=\Omega
\cos{\theta_{\text{Lat}}}$ and $\Omega_z=\Omega
\sin{\theta_{\text{Lat}}}$ with $\Omega=7.27\times
10^{-5}~\text{rad}/\text{s}$ and $\theta_{\text{Lat}}=37.4$
degrees North latitude. The initial position of the atom in the
lab is taken as $\v{r}(0)=(x_0,y_0,0)$, with $x_0=1~\text{mm}$ and
$y_0=1~\text{mm}$.  The initial velocity is $\v{\dot{r}}(0)=(v_x,
v_y, v_z)$, with $v_x = 1~\text{mm}/\text{s}$, $v_y =
1~\text{mm}/\text{s}$, and $v_z = 13.2~\text{m}/\text{s}$.}
\end{center}
\end{table}

Using the above method, we computed the phase shift response for a
single atom interferometer, and the results are shown in Table
\ref{Tab:full phases}.  The values of the experimental parameters
used to generate this list are representative of the
$8.8~\text{m}$ apparatus described previously.  Many of these
terms are common to both species, and in order obtain our
$<10^{-15}g$ sensitivity, we rely on their common mode
cancellation.  In Table \ref{Tab:EP phases} we compute the
differential phase shift between a $^{87}$\!Rb and a $^{85}$\!Rb
interferometer.  The two species have different masses $m$ and
second order Zeeman coefficients $\alpha$, as well as potentially
different launch kinematics $\v{r}(0)$ and $\v{\dot{r}}(0)$.  To
create Table \ref{Tab:EP phases}, we parameterized the launch
kinematics with a differential velocity $\v{\delta v}=(\delta v_x,
\delta v_y, \delta v_z)$ and initial position $\v{\delta
r}=(\delta x, \delta y, \delta z)$ between the centroids of the
two isotope clouds.  Residual systematic phase errors are the
result of differential accelerations that arise from gravity
gradients, second gravity gradients, coriolis and centrifugal
forces, and magnetic forces on the atoms.

\begin{table}
\begin{center}
\begin{math}
\begin{array}{lccc}
\hline\\
\qquad\qquad & \qquad\qquad \text{Phase shift}\qquad\qquad  & \qquad \text{Size (rad)}\qquad & \qquad\text{Fractional size}\qquad \\
\hline\hline\\
 1 & -\frac{1}{2} \left(\frac{1}{\text{m85}}-\frac{1}{m_{87}}\right) \hbar  k_{\text{eff}}^2 T_{zz} T^3 &
   1.66\times 10^{\text{-2}} & 5.83\times 10^{\text{-11}} \\
 2 & 2 k_{\text{eff}} \delta v_x \Omega_y T^2 & 3.35\times 10^{\text{-3}} & 1.18\times
   10^{\text{-11}} \\
 3 & -k_{\text{eff}} T_{zz} \delta v_z T^3 & 1.44\times 10^{\text{-4}} & 5.05\times 10^{\text{-12}} \\
 4 & -\frac{3}{2} \left(\frac{1}{m_{85}}-\frac{1}{m_{87}}\right) \hbar
   k_{\text{eff}}^2 \Omega_y^2 T^3 & -5.40\times 10^{\text{-5}} & 1.90\times 10^{\text{-13}} \\
 5 & -3 k_{\text{eff}} \Omega_y^2 \delta v_z  T^3 & -4.68\times 10^{\text{-6}} & 1.64\times 10^{\text{-14}} \\
 6 & -k_{\text{eff}} T_{zz} \delta z T^2 & 8.93\times 10^{\text{-7}} & 3.14\times 10^{\text{-15}} \\
 7 & -k_{\text{eff}} \delta y \Omega_y \Omega_z T^2 & -7.41\times 10^{\text{-7}} &
   2.60\times 10^{\text{-15}} \\
 8 & 3 k_{\text{eff}} \delta v_y \Omega_y \Omega_z T^3 & 2.98\times 10^{\text{-7}} &
   1.05\times 10^{\text{-15}} \\
 9 & -\frac{7}{12} \left(\frac{1}{m_{85}}-\frac{1}{m_{87}}\right) \hbar
   k_{\text{eff}}^2 Q_{zzz} v_z T^4 & -1.61\times 10^{\text{-7}} & 5.65\times 10^{\text{-16}} \\
 10 & \frac{3}{8} \left(\frac{1}{m_{85}}-\frac{1}{m_{87}}\right) \hbar k_{\text{eff}}^2 g Q_{zzz} T^5
   & 1.03\times 10^{\text{-7}} & 3.63\times 10^{\text{-16}} \\
 11 & -\left(\frac{\alpha_{85}}{m_{85}}-\frac{\alpha_{87}}{m_{87}}\right)
   \hbar  k_{\text{eff}} B_0 (\partial_z B) T^2 & -9.94\times 10^{\text{-8}} & 3.49\times 10^{\text{-16}} \\
 12 & -2 k_{\text{eff}} T_{xx} \delta x \Omega_y T^3 & -6.92\times 10^{\text{-8}} &
   2.43\times 10^{\text{-16}}\\
\hline
\end{array}
\end{math}
\caption{\label{Tab:EP phases} Differential phase shift between
$^{87}$\!Rb and $^{85}$\!Rb.  To create the differential phase
shift list we parameterized the launch kinematics with a
differential velocity $(\delta v_x = 1~\mu \text{m}/\text{s},
\delta v_y = 1~\mu \text{m}/\text{s}, \delta v_z =
12~\mu\text{m}/\text{s})$ and position $(\delta x = 1~\mu
\text{m}, \delta y = 1~\mu \text{m}, \delta z = 10~\text{nm})$
between the centroids of the two isotope clouds.  All other
parameters are the same as in Table \ref{Tab:full phases}.  Column
3 shows the fractional size of each term compared to the
acceleration signal $k_{\text{eff}} g T^2$.  We include all terms
with a fractional phase shift $> 10^{-16}$.}
\end{center}
\end{table}

As justified below, we expect to achieve experimental parameters
that reduce the majority of the systematic errors below our
experimental threshold.  However, the first several terms in Table
\ref{Tab:EP phases} are still too large.  In order to further
reduce these backgrounds, we can employ propagation reversal to
suppress all terms $\propto \keff^2$.   This well--known technique
entails reversing the laser propagation vector
$\v{\keff}\longrightarrow-\v{\keff}$ on subsequent trials and then
subtracting the two results \cite{Fixler}.  This suppresses terms
1, 4, 9, and 10 by $\Delta \keff/\keff$, where $\Delta \keff$ is
the error in $\keff$ made as a result of the reversal.  Reducing
these terms below our systematic threshold requires $\Delta
\keff/\keff<10^{-5}$.  The main acceleration signal and all other
terms linear in $\keff$ are not suppressed by this subtraction.

After propagation vector reversal, the last important background
phase shifts arise from the differential coriolis and centrifugal
acceleration between the isotopes (Table \ref{Tab:EP phases} terms
2, 5, 7, and 8), and from the Earth's gravity gradient (Table
\ref{Tab:EP phases} terms 3 and 6).  We discuss the techniques
used to control these remaining systematics in Section
\ref{Sec:control systematics}.

\subsubsection{Gravity Inhomogeneities}

\label{Sec:Gravity Inhomogeneities}

The Taylor series expansion of the gravitational potential (see
Eq. \ref{Eq:GravitationalTaylorSeries}) is a good approximation of
the coarse structure of Earth's gravity on length scales of $\sim
R_e$, the radius of the Earth.  However, local gravity can also
vary on much shorter length scales in a way that depends on the
specific mass distribution surrounding the experiment, and these
gravity inhomogeneities can result in spurious phase shifts. Since
these inhomogeneities can be rapidly spatially varying, the Taylor
series expansion is not well-suited for their description.
Instead, we leverage the fact that these inhomogeneities are
typically small in magnitude and solve for the induced phase shift
using first--order perturbation theory \cite{cct}.  This
linearization allows us to make a Fourier decomposition of the
phase shift response in terms of the spatial wavelengths of the
local $g$--field.

First, we assume a one-dimensional gravitational potential
perturbation of the form $\delta\phi(z)$.  The gravity field
perturbation along the vertical ($z$) direction is defined as
$\delta g_z(z)\equiv - \partial_z\delta\phi$ and may be written as
\be\delta g_z(z) = \int  \widetilde{\delta g}_z(\lambda)
e^{\frac{i 2 \pi z}{\lambda}} d \lambda\ee where
$\widetilde{\delta g}_z(\lambda)$ is the Fourier component of a
gravity perturbation with wavelength $\lambda$.  The total phase
shift due to gravity inhomogeneities summed over all wavelengths
is \be\Delta\phi_{g}=\int T_{gz}(\lambda) \widetilde{\delta
g}_z(\lambda) d\lambda\ee where $T_{gz}(\lambda)$ is the
interferometer's gravity perturbation response function.
Qualitatively, the response to short wavelengths is suppressed
since the interferometer averages over variations that are smaller
than its length \cite{bigGR}.  The response is flat for
wavelengths longer than the scale of the interferometer, and in
the limit where $\lambda\sim R_e$ this analysis smoothly
approaches the results of our Taylor series calculation described
above.

\begin{figure}
\begin{center}
\includegraphics[width=350pt]{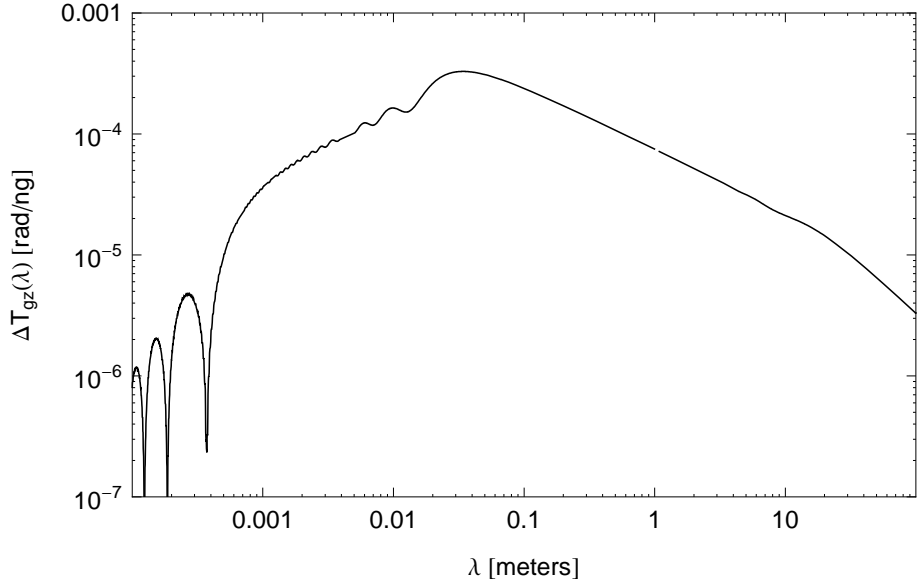}
\caption{\label{Fig:gravity wrinkles} Differential gravity
response function versus spatial wavelength $\lambda$ between
simultaneous $^{87}$\!Rb and $^{85}$\!Rb interferometers.  Short
wavelengths are averaged over by each individual interferometer,
while long wavelength inhomogeneities cancel as a common-mode
between the two species.  This response curve assumes identical
launch kinematics for the two isotopes.}
\end{center}
\end{figure}

For the $^{87}$\!Rb--$^{85}$\!Rb EP measurement, we are interested
in the differential phase response between the isotopes.   Figure
\ref{Fig:gravity wrinkles} shows the differential response
function $\Delta T_{gz}(\lambda)\equiv
\left|(T_{gz})_{87}-(T_{gz})_{85}\right|$ for gravity
inhomogeneities.  Once again, short wavelength variations are
suppressed since each interferometer spatially averages over a
$\sim 10~\text{m}$ region.  The peak response occurs at a length
scale set by the spatial separation of the arms of a single
interferometer $\Delta z=\frac{\hbar\keff}{m}T\sim 16~\text{mm}$.
Perfect differential cancellation between isotopes is not achieved
because the spatial separation of the arms is mass dependent.
Additionally, the long wavelength differential response is
suppressed because the differences between the isotope
trajectories are negligible when compared to variations with
length scales much longer than $\Delta z$.

The differential response curve allows us to compute systematic
errors arising from the specific gravity environment of our
interferometer.  Quantitative estimates of these effects requires
knowledge of the local $\delta g_z(z)$, which may be obtained
through a combination of modelling and characterization.  The atom
interferometer itself can be used as a precision gravimeter for
mapping $\delta g_z(z)$ in situ.  By varying the launch velocity,
initial vertical position, and interrogation time $T$, the
position of each gravity measurement can be controlled.

\begin{figure}
\begin{center}
\includegraphics[width=350pt]{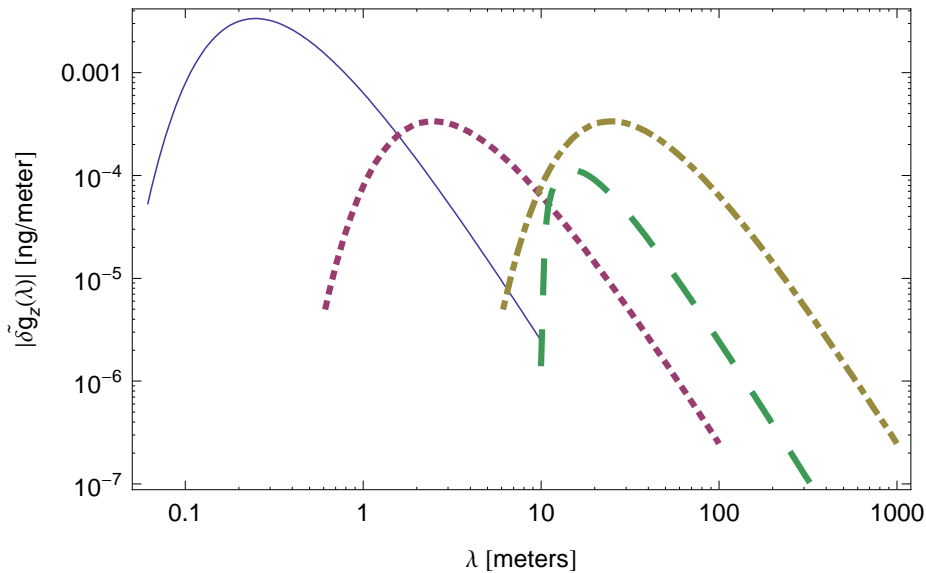}
\caption{\label{Fig:gravity sources} The magnitude power spectra
of the local gravitational field, $\widetilde{\delta g}_z$, for
several example sources. The solid (blue) curve is a
$10^{-2}~\text{kg}$ point source, 10 cm from the center of the
interferometer. Similarly, the dotted (purple) curve is a 1 kg
source at 1 m and the dash-dotted (yellow) curve is 1000 kg at 10
m. The long-dashed (green) curve is a thin 10 m long rod of mass
10 kg, parallel to the interferometer, whose center is 1 m from
the interferometer.}
\end{center}
\end{figure}

Figure \ref{Fig:gravity wrinkles} shows that the differential
$^{87}$\!Rb--$^{85}$\!Rb interferometer is maximally sensitive to
short wavelength ($\lambda \sim 1-10~\text{cm}$) gravitational
inhomogeneities. To investigate the impact of local uneven mass
distributions on the experiment, we compute the spectrum
$\widetilde{\delta g}_z(\lambda)$ of various sources at different
distances from the interferometer.  These results are shown in
Fig. \ref{Fig:gravity sources}. When combined with our response
function (Fig. \ref{Fig:gravity wrinkles}), we see that for
typical mass inhomogeneities, only those that are within a few
centimeters of the interferometer can cause potentially
significant systematic phase shifts.  These nearby inhomogeneities
result in phase errors of $\sim 10^{-6}~\text{rad}$, which is
slightly above our target sensitivity. It will therefore be
especially important for the EP measurement that we characterize
the local $g$--field at the centimeter scale.

\subsubsection{Magnetic field inhomogeneities}
\label{Sec:Magnetic Inhomgeneities} The linear expansion of
$\v{B}$ in Eq. \ref{Eq:TaylorMagneticFields} approximates large
scale variation of the magnetic field. However, local field
inhomogeneities may exist on short length scales due to the
presence of nearby magnetic materials. These variations are not
well approximated by a Taylor series expansion.  Using the same
procedure described above for gravity inhomogeneities, we write
the local magnetic field as \be B(z)=\int \widetilde{B}_z(\lambda)
e^{\frac{i 2 \pi z}{\lambda}} d \lambda\ee where
$\widetilde{B}_z(\lambda)$ is the Fourier component of a field
perturbation with wavelength $\lambda$.  The total phase shift
from magnetic field inhomogeneities is \be \Delta\phi_{B}=\int
T_{Bz}(\lambda) \widetilde{B}_z(\lambda) d\lambda \ee Here
$T_{Bz}(\lambda)$ is the interferometer's magnetic inhomogeneity
response function.

\begin{figure}
\begin{center}
\includegraphics[width=350pt]{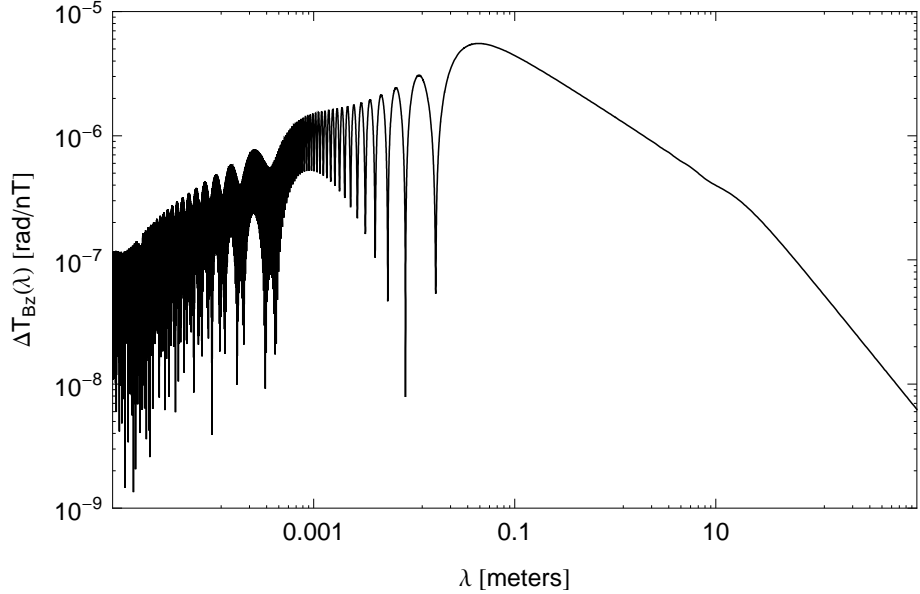}
\caption{\label{Fig:magnetic_field_wrinkles} Differential magnetic
field response function between simultaneous $^{87}$\!Rb and
$^{85}$\!Rb interferometers.  Short wavelengths are averaged over
by each individual interferometer, while long wavelength
inhomogeneities cancel as a common-mode between the two species.
This response curve assumes identical launch kinematics for the
two isotopes.}
\end{center}
\end{figure}

As with gravity above, we compute the differential response
function $\Delta T_{Bz}(\lambda)\equiv
\left|(T_{Bz})_{87}-(T_{Bz})_{85}\right|$ between $^{87}$\!Rb and
$^{85}$\!Rb (see Fig. \ref{Fig:magnetic_field_wrinkles}).  The
differential response arises because the isotopes have different
second order Zeeman coefficients $\alpha$, as well as different
masses.  This sensitivity curve drives our magnetic shield design
requirements, as discussed in Section \ref{Sec:control magnetic
fields}.

\subsection{Controlling potential systematic errors} \label{Sec:control
systematics}

\subsubsection{Rotation of the Earth}

The largest systematic term in the phase shift expansion for a
dual species differential interferometer after propagation
reversal is due to the rotation of the Earth.  Specifically, a
differential acceleration due to the coriolis force occurs if the
isotopes have different transverse velocities $\delta v_x$ (Table
\ref{Tab:EP phases}, term 2).  Reducing this phase shift below our
systematic threshold would require $\delta
v_x<10^{-11}~\text{m}/\text{s}$, which is challenging.  However,
this specification can be relaxed by artificially making the
rotation rate zero.

To a good approximation\footnote{The atoms are also weakly coupled
electromagnetically and gravitationally to the local environment,
which is fixed to the rotating Earth.  These cross-couplings to
rotation are generally not important because the dominant
gravitational interaction with the Earth is spherically symmetric,
and all electromagnetic interactions with the atom (e.g. with the
applied magnetic bias field) are naturally small.}, the atoms are
only affected by the Earth's rotation through their coupling to
the laser, so the coriolis acceleration can be eliminated by
rotating the laser in the opposite direction of the Earth's
rotation.  In order to calculate the effect of this rotation
compensation, we performed the phase shift calculation using a
rotating $\v \keff$. Following the work of \cite{AchimMetrologia},
we use a retro-reflection configuration to deliver the laser beams
$\v k_1$ and $\v k_2$ to the atoms. We rotate $\v \keff$ by
actuating the retro-reflection mirror. As a result, the incoming
beam remains pointing along the $z$-direction and only the
reflected beam rotates.  With this configuration, $\v \keff$ is
given by \be \v \keff = -2 k \v{\hat{n}} (\v{\hat{n}} \cdot
\v{\hat{k}})\ee where $\v{\hat{n}}$ is  the time-dependent unit
normal vector of the retro-reflection mirror, and $\v{\hat{k}}$ is
a unit vector in the direction of the fixed delivery beam. Notice
that the direction of $\v \keff$ rotates as desired, but its
length now depends on angle\footnote{This small change in
magnitude of $\keff$ does not lead to any problematic phase errors
in the interferometer since the total angle through which the
laser rotates is only $\Delta\theta=2\Omega_{\text{Earth}} T \sim
10^{-4}~\text{rad}$, and the effect is
$\mathcal{O}(\Delta\theta^2)$.}.

The resulting phase shift list appears in Table \ref{Tab:EP rot
comp phases}, with $\delta\Omega_y$ and $\delta\Omega_z$ the
errors in the applied counter-rotation rate.  Assuming a
transverse velocity difference of $\delta v_x\sim
1~\mu\text{m}/\text{s}$, these rotation compensation errors must
be kept below $10^{-5}\Omega_{\text{Earth}}\approx
1~\text{nrad}/\text{s}$. Methods for measuring angles with
nanoradian precision have already been demonstrated
\cite{optical_lever}. In order to actuate the mirror at this level
of precision we can use commercially available sub-nm accurate
piezo-electric actuators along with active feedback.

Notice that not all rotation-related phase errors are removed by
rotation compensation.  Terms that arise from the differential
centrifugal acceleration between the isotopes (e.g., Table
\ref{Tab:EP rot comp phases} terms 3 and 4) are not suppressed.
Physically, this is a consequence of the fact that the
retro-reflection mirror that we use to change the laser's angle is
displaced from the center of rotation of the Earth by $R_e$.
Therefore, although we can compensate for the angle of the laser
by counter-rotating, the retro-reflection mirror remains attached
to the rotating Earth, leading to a centrifugal acceleration of
the phase fronts.  After propagation reversal, the only term of
this type that is significant is $\sim k_{\text{eff}} \Omega_y^2
\delta v_z T^3$ (Table \ref{Tab:EP rot comp phases} term 4).
However, this term is smaller than and has the same scaling with
experimental control parameters as the gravity gradient phase
shift (Table \ref{Tab:EP rot comp phases} term 2), so the
constraints described in Section \ref{Sec:Gravity Gradients} to
suppress the gravity gradient terms are sufficient to control this
centrifugal term as well.

One potential obstacle in achieving the required transverse
velocity constraint of $\delta v_x\sim 1~\mu\text{m}/\text{s}$ is
the expected micro-motion the atoms experience in the TOP magnetic
trap prior to launch \cite{TOPmicromotion}.  Micro-motion orbital
velocities in a tight TOP trap such as ours can approach $\sim
1~\text{cm}/\text{s}$ in the transverse plane.  Although the
differential orbital velocities are suppressed by the
$^{87}$\!Rb--$^{85}$\!Rb mass ratio, the resulting $\delta v_x\sim
100~\mu\text{m}/\text{s}$ is still too large.  This problem can
potentially be solved by adiabatically reducing the magnetic field
gradient and increasing the rotating field frequency prior to
launch.

\begin{table}
\begin{center}
\begin{math}
\begin{array}{lccc}
\hline\\
\qquad\qquad & \qquad\qquad \text{Phase shift}\qquad\qquad  & \qquad \text{Size (rad)}\qquad & \qquad\text{Fractional size}\qquad \\
\hline\hline\\
1 & -\frac{1}{2}
\left(\frac{1}{m_{85}}-\frac{1}{m_{87}}\right)\hbar
k_{\text{eff}}^2 T_{zz} T^3 &
   1.66\times 10^{\text{-2}} & 5.83\times 10^{\text{-11}} \\
 2 & -k_{\text{eff}} T_{zz} \delta v_z T^3 & 1.44\times 10^{\text{-4}} & 5.05\times 10^{\text{-12}} \\
 3 & -\frac{3}{2} \left(\frac{1}{m_{85}}-\frac{1}{m_{87}}\right)\hbar k_{\text{eff}}^2 \Omega_y^2 T^3
   & -5.40\times 10^{\text{-5}} & 1.90\times 10^{\text{-13}} \\
 4 & -3 k_{\text{eff}} \Omega_y^2 \delta v_z T^3 & -4.68\times 10^{\text{-6}} & 1.64\times 10^{\text{-14}} \\
 5 & -k_{\text{eff}} T_{zz} \delta z T^2 & 8.93\times 10^{\text{-7}} & 3.14\times 10^{\text{-15}} \\
 6 & -\frac{7}{12} \left(\frac{1}{m_{85}}-\frac{1}{m_{87}}\right)\hbar k_{\text{eff}}^2 Q_{zzz} v_z
   T^4 & -1.61\times 10^{\text{-7}} & 5.65\times 10^{\text{-16}} \\
 7 & \frac{3}{8} \left(\frac{1}{m_{85}}-\frac{1}{m_{87}}\right)\hbar k_{\text{eff}}^2 Q_{zzz} g T^5 &
   1.03\times 10^{\text{-7}} & 3.63\times 10^{\text{-16}} \\
 8 & -\left(\frac{\alpha_{85}}{m_{85}}-\frac{\alpha_{87}}{m_{87}}\right)\hbar k_{\text{eff}} B_0
   (\partial_z B) T^2 & -9.94\times 10^{\text{-8}} & 3.49\times 10^{\text{-16}} \\
 9 & k_{\text{eff}} T_{xx} \delta x \Omega_y T^3 & 3.46\times 10^{\text{-8}} & 1.22\times
   10^{\text{-16}} \\
 10 & -2  k_{\text{eff}} \delta v_x \delta\Omega_y T^2 & -3.35\times 10^{\text{-8}} & 1.18\times
   10^{\text{-16}} \\
\hline
\end{array}
\end{math}
\caption{\label{Tab:EP rot comp phases} Differential phase shift
list with rotation compensation.  Terms 1, 3, 6, and 7 will be
suppressed by the propagation reversal technique described in
Section \ref{Sec:EP phase shift calc}.}
\end{center}
\end{table}

\subsubsection{Gravity gradients}
\label{Sec:Gravity Gradients} The largest systematic background
after rotation compensation is due to the gravity gradient along
the vertical direction of the apparatus ($T_{zz}=\partial_z g_z$).
Since gravity is not uniform, the two isotopes experience a
different average acceleration if their trajectories are not
identical.  This effect causes a differential phase shift
proportional to the initial spatial separation and initial
velocity difference between the isotopes (see Table \ref{Tab:EP
rot comp phases}, terms 2 and 5).  Assuming a spherical Earth
model, the gravity gradient felt by the atoms is $T_{zz}\sim
3\times 10^{-16}g/\text{nm}$, which means that the initial
vertical position difference between the isotopes $\delta z$ must
be $< 1~\text{nm}$ and the initial vertical velocity difference
$\delta v_z$ must be $< 1~\text{nm}/\text{s}$ in order to reduce
the systematic phase shift beneath our threshold.

The experiment is designed to initially co-locate the two isotope
clouds at the nm level by evaporative cooling both species in the
same magnetic trap.  For trapping, we use the state
$\ket{F=2,m_F=2}$ for $^{87}$\!Rb and $\ket{F=3,m_F=3}$ for
$^{85}$\!Rb since these states have the same magnetic moment
\cite{bloch 85 87 cooling}.  The mass difference between the
isotopes leads to a differential trap offset in the combined
magnetic and gravitational potential given by \be\Delta
z_{\text{trap}}=\frac{g\Delta m}{\mu_B B''}\ee where $\Delta m$ is
the $^{87}$\!Rb--$^{85}$\!Rb mass difference, $B''$ is the
magnetic field curvature of the trap, and $\mu_B$ is the Bohr
magneton. Our TOP magnetic trap is designed to provide a field
curvature $B''\sim 4\times 10^5~\text{Gauss}/\text{cm}^2$ which
reduces the trap offset to $\Delta z_{\text{trap}}\approx
10~\text{nm}$.  The resulting systematic error is $\sim
10^{-14}g$, but it can be subtracted during the analysis given a
knowledge of $\Delta z_{\text{trap}}$ at the $\sim 10\%$ level.
This offset can be inferred from a measurement of the field
curvature $B''$ of the trap (e.g., by measuring the trap
oscillation frequency).  The gravity gradient must also be known,
but this can be characterized in situ by using the interferometer
as a gradiometer \cite{Fixler}.

Control of the gravity gradient phase shift (Table \ref{Tab:EP rot
comp phases} term 2) requires that the differential launch
velocity be $\delta v_z\leq 1~\text{nm}/\text{s}$.  Therefore we
cannot employ standard launch techniques (e.g., moving molasses)
since the velocity uncertainty is fundamentally limited by the
photon recoil velocity $v_R\sim 6~\text{mm}/\text{s}$ due to
spontaneous emission.  Instead, the atoms are launched using an
accelerated optical lattice potential \cite{Phillips2002:JPhysB}.
We launch the two species using the same far-detuned ($\sim 200
\text{GHz}$) optical lattice, coherently transferring $\sim 2200
\hbar k$ of momentum to each cloud. Because the two species have
different masses, they have different Bloch oscillation times
$\tau_B=\frac{\hbar\keff}{m a}$, where $a$ is the lattice
acceleration.  As a result, after adiabatically ramping down the
lattice potential, the two species are in different momentum
eigenstates since they have absorbed a different number of
photons.  The differential velocity after launch is then \be
\delta
v_L=\hbar\keff\left(\frac{N_{85}}{m_{85}}-\frac{N_{87}}{m_{87}}\right)\ee
where $N_{85}$ and $N_{87}$ are the number of photons transferred
to $^{85}$\!Rb and $^{87}$\!Rb, respectively.  We choose the
integers $N_{85}=2168$ and $N_{87}=2219$ such that their ratio is
as close to the isotope mass ratio as possible, resulting in
$\delta v_L\sim 12~\mu \text{m}/\text{s}$.  After launch, we can
perform a velocity selective transition to pick out a common class
from the overlapping distributions of the two isotopes, which at
the expense of atom number could conceivably allow us to achieve
our differential velocity constraint.

There are several additional ways to reduce the gravity gradient
systematics beyond precise control of the launch kinematics.  We
can implement a 4-pulse sequence
($\frac{\pi}{2}-\pi-\pi-\frac{\pi}{2}$) which suppresses all phase
shift terms $\propto T^3$ at the cost of an order one loss in
acceleration sensitivity \cite{boris}.  This eliminates the
velocity dependent gravity gradient phase shifts but would still
require that we maintain tight control over the initial
differential vertical position between the isotope clouds.
Secondly, we can potentially reduce the local gravity gradient
$T_{zz}$ by applying appropriate trim masses around the apparatus.
It has been shown \cite{bigGR} that in principle a local mass
distribution can effectively cancel the gravity gradient of the
Earth for a $10~\text{m}$-scale apparatus.  Reducing $T_{zz}$ by
an order of magnitude would relax our initial position constraint
to the level provided by the expected value of $\Delta
z_{\text{trap}}$, thereby removing the requirement for subtraction
during data analysis.

\subsubsection{Magnetic fields}
\label{Sec:control magnetic fields}

The magnetic field phase shift appearing in Table \ref{Tab:EP rot
comp phases} (term 8) constrains the maximum allowed linear field
gradient to $\partial_z B < 0.1~\text{nT}/\text{meter}$. In the
interferometer region, the measured gradient of the Earth's field
is $\sim 3~\mu\text{T}/\text{m}$, and therefore we require a
shielding ratio of at least $\sim 5\times 10^4$.  In addition to
suppressing the field gradient, the magnetic response function
(see Fig. \ref{Fig:magnetic_field_wrinkles}) indicates that the
field must be uniform on length scales $\sim 1~\text{cm}$.  Large
magnetic shields with similar performance have been demonstrated
\cite{NeutronShield}.  The magnetic shielding for our
interferometer region is provided by a three--layer concentric
cylindrical shield made of high permeability material.  To
maintain a pristine magnetic environment, we use an aluminum
vacuum chamber and non-magnetic materials inside the shielded
region.

In order to verify the performance of the magnetic shield, we must
characterize the field.  As with gravity inhomogeneities, the atom
interferometer can be used to map the local magnetic field in
situ, in this case by using a magnetic field sensitive ($m_F \neq
0$) state \cite{characterizeBwithAtoms}.

\section{Conclusion}
In these notes we have given an overview of the light-pulse
method, and discussed applications to inertial navigation and a
test of the Equivalence Principle.  As the field continues to
progress, we see two trends evolving.  First, a steady evolution
in the technology associated with laser manipulation of atoms will
lead to progressive development of compact, field-ready inertial
sensors.  Such developments include integrated photonics packages
for the laser and optics paths, and integrated optics bench and
vacuum systems.  On the other hand, continued evolution of high
performance science instruments will extend the physics reach of
this technology.  For example, next generation light-pulse atom
interference systems appear capable of placing superb limits on
atom charge neutrality \cite{atomneutrality}, making terrestrial
tests of General Relativity \cite{bigGR,GR}, and possibly
detecting low frequency gravitational waves
\cite{gravitywaves,BigGravitywaves}.

\end{document}